\documentclass[twoside]{article}

\usepackage{PRIMEarxiv}
\usepackage[utf8]{inputenc} 
\usepackage[T1]{fontenc}    
\usepackage{hyperref}       
\usepackage{url}            
\usepackage{booktabs}       
\usepackage{amsfonts}       
\usepackage{nicefrac}       
\usepackage{microtype}      
\usepackage{lipsum}
\usepackage{fancyhdr}       
\usepackage{graphicx}       
\graphicspath{{media/}}     
\usepackage{import}
\usepackage{caption}
\usepackage{subcaption}
\usepackage{array}
\usepackage{multirow}
\usepackage{rotating}
\usepackage{float}
\usepackage{MnSymbol}
\usepackage{lscape}
\usepackage{longtable}
\usepackage{ragged2e}
\usepackage{graphicx}
\usepackage{tabularx}
\usepackage{booktabs}
\usepackage{hyperref}
\usepackage[numbers]{natbib}
\usepackage{soul} 

\title{Human-Centered Explainability in Interactive Information Systems: A Survey}

\author{
  Yuhao Zhang \\
  Peking University \\
  China \\
  The University of Oklahoma \\
  USA \\
  \texttt{zyh\_ruc@163.com}  \\
   \And
  Jiaxin An \\
  The University of Texas at Austin \\
  USA \\
  \texttt{jiaxin.an@utexas.edu} \\
  \AND
  Ben Wang \\
  The University of Oklahoma \\
  USA \\
  \texttt{benw@ou.edu}  \\
  \And
  Yan Zhang \\
  The University of Texas at Austin \\
  USA \\
 \texttt{yanz@utexas.edu} \\
 \And
  Jiqun Liu \thanks{Corresponding author: jiqunliu@ou.edu}\\
  The University of Oklahoma \\
  USA \\
  \texttt{jiqunliu@ou.edu} \\
}

\begin{document}
\maketitle

\begin{abstract}
\textit{Human-centered explainability} has become a critical foundation for the responsible development of interactive information systems, where users must be able to understand, interpret, and scrutinize AI-driven outputs to make informed decisions. This systematic survey of literature aims to characterize recent progress in user studies on explainability in interactive information systems by reviewing how explainability has been conceptualized, designed, and evaluated in practice. Following PRISMA guidelines, eight academic databases were searched, and 100 relevant articles were identified. A structural encoding approach was then utilized to extract and synthesize insights from these articles. The main contributions include 1) five dimensions that researchers have used to conceptualize explainability; 2) a classification scheme of explanation designs; 3) a categorization of explainability measurements into six user-centered dimensions. The review concludes by reflecting on ongoing challenges and providing recommendations for future exploration of related issues. The findings shed light on the theoretical foundations of human-centered explainability, informing the design of interactive information systems that better align with diverse user needs and promoting the development of systems that are transparent, trustworthy, and accountable.
\end{abstract}


\section{Introduction}
In recent years, Artificial Intelligence (AI) has been integrated into and evaluated within a wide range of interactive information systems (e.g., search and recommendation \cite{zhang2020explainable, cheng2019explaining, ai2019explainable, sun2021unsupervised, chen_reference-dependent_2023, xu2023reusable, li2023personalized}) across various domains, such as healthcare (e.g., \cite{payrovnaziri2020explainable,holzinger2019causability}), finance (e.g., \cite{yeo2025comprehensive}), training and entertainment (e.g., \cite{zhao2019personalized,lee2018explainable, wang2021personalized}), to name a few. These systems increasingly deploy complex, high-dimensional AI algorithms, which notoriously operate as “black-boxes.” This opacity poses significant challenges, including those related to trust, accountability, and fairness. Consequently, there is a growing demand for humans to pursue more controllable processes, and efforts to encourage algorithms to explain their outputs are being increasingly prioritized in practice. Building on these advancements, explainable AI (XAI) arises, with explainability emerging as a vital property of AI systems that aim at making algorithmic reasoning processes more transparent and trustworthy for human users \cite{burkart2021survey, liu2022trustworthy}.

Although existing research in XAI has primarily focused on developing algorithmic techniques to enhance technical explainability, there has been a growing emphasis in recent years on evaluating the effectiveness of these explanations. In the early stages of such research, humans are less involved in the evaluation process and receive less attention, which results in users being unable to understand and utilize the explanations sufficiently \cite{narayanan2018humans,miller2019explanation}. The effectiveness of explanations lies in the perception and reception of the person at the receiving end \cite{liao2021human}. Therefore, embracing more \textit{human-centered} approaches and getting humans involved in the explainable AI development and research are gaining momentum~\cite{nguyen2022towards}. Many studies, particularly those from the Human-Computer Interaction (HCI) community, have explored human-centered approaches to further enhance AI explainability~\cite{ehsan_who_2021, smith2020no}. Such studies have targeted various stakeholders, from laypeople to experts \cite{cheng2019explaining,ehsan_who_2021, panigrahi2025interactivity}. For example, \cite{ehsan_who_2021} conducted a mixed-methods study to examine how users’ AI backgrounds shape their perceptions of AI explanations, revealing disparities that inform the design of more inclusive and trustworthy explainable AI systems. 

Building on these advancements, scholars seek to gain overviews of the ever-growing field to build up knowledge of Human-Centered Explainable AI. For instance, \cite{nguyen_how_2024} has examined explainable interfaces in 53 publications and mapped out the current state of user interface and user experience design aspects of XAI. The work by \cite{haque_explainable_2023} included 58 articles and identified the dimensions of end users’ explanation needs with an aim to distill the effects of explanations on end users’ perceptions.

Despite these promising developments, three critical issues persist. First, while existing studies mainly focus on AI, they overlook the practice of explainability in traditional information systems, such as the explanations provided by search engines using classic models and phishing reminders. These practices can still be valuable for today's AI-assisted systems. Second, although human-centered approaches have gained considerable traction, a comprehensive synthesis of empirical user studies that directly examine users’ perceptions, behaviors, and experiences with explainable systems remains lacking. Third, while recent reviews have made valuable contributions, there have not been systematic efforts to offer an integrated perspective that connects definitions, design practices, and evaluation approaches within interactive information systems. To address these gaps, this review systematically examines the current state of user studies on explainability in interactive information systems, aiming to clarify how explainability has been conceptualized, designed, and evaluated in practice. Accordingly, we seek to answer the following research questions (RQs):

\textbf{RQ1}: How is explainability defined in research on interactive information systems? 

\textbf{RQ2}: How are explainability features integrated into the interface design and user interactions of interactive information systems?

\textbf{RQ3}: How is the explainability of interactive information systems evaluated? 

This review employs the Preferred Reporting Items for Systematic Reviews and Meta-Analyses (PRISMA) guideline \cite{page_prisma_2021}, a standard framework for conducting transparent and rigorous literature reviews. Its application enables us to systematically identify and analyze 100 publications while maintaining methodological clarity and consistency. The remainder of the paper is organized as follows. Section 2 introduces related work in the field and situates our study within the existing literature. Section 3 describes the methods used for identifying, screening, and analyzing the literature. Section 4 presents the findings in response to the three research questions above. Finally, Section 5 discusses the implications of our findings and outlines directions for future research. 

\section{Related work}\label{related_work}
\subsection{The definition of explainability}
\textit{Explainability} is a complex and multifaceted concept characterized by varying definitions and terminologies across different research domains. Terms such as interpretability, understandability, and transparency are frequently employed interchangeably or synonymously with explainability, despite notable overlaps and subtle distinctions among them \cite{barredo_arrieta_explainable_2020, vilone_notions_2021}. Different disciplines prioritize distinct aspects and dimensions of explainability, reflecting the diverse objectives and methodological concerns inherent to each domain.

Despite diverse definitions, previous research has attempted systematically categorize terminologies of explainability based on different dimensions. \cite{mohseni_multidisciplinary_2021} categorized and summarized the concepts of XAI through aspects including scope, approach, goal, and presentation format. Scope distinguishes between global and local explanations, depending on whether the explanation focuses on the entire model or specific results \cite{lundberg_local_2020}. The approach focuses on the model itself, contrasting inherently interpretable models with ad-hoc explainers designed for interpreting black-box models \cite{lundberg_unified_2017}. The goal refers to the objectives or types of user questions addressed by explanations, such as "How,"\cite{lombrozo_explanation_2009} "Why,"\cite{ribeiro_why_2016} "Why-not,"\cite{vermeulen_pervasivecrystal_2010} "What-if,"\cite{kocielnik_will_2019} "How-to,"\cite{lim_why_2009} and "What-else"\cite{cai_human-centered_2019} explanations. Lastly, Presentation format is how explanations are conveyed, including visual forms (e.g., saliency maps \cite{simonyan_deep_2014}), verbal descriptions (e.g., natural language explanations \cite{lakkaraju_interpretable_2016}), explainable interfaces with multiple modalities \cite{myers_answering_2006}, and analytic presentations (e.g., numerical metrics or structured visualizations \cite{hohman_visual_2019}). This dimension also includes considerations of content specifics and language.

To enrich the conceptual clarity of XAI terminology, \cite{bellucci_towards_2021} proposed a taxonomy of key terms from the perspective of Explainable Information Systems (XIS), including general, explanation-specific, and system-specific terminologies. General terminologies include concepts fundamental to the construction and evaluation of XIS, such as user, explanation, explainability, interpretability, and trust. Within this general perspective, the user with different background can influence the effectiveness of explanations \cite{ribeiro_why_2016}. Explanation is defined as the output of an XIS, interpretability and explainability represent the capability of an XIS to produce understandable and meaningful outputs, and trust ensures that explanations are effective, credible, and acceptable from the user's perspective \cite{gunning_darpas_2019}.

Explanation-specific terminology comprises four primary elements: focus, means, modality, and reasoning. Focus aligns with the concept of scope in [93], distinguishing between global and local explanations. Means differentiates direct explanations from post hoc explanations, which is relevant to approach. Modality addresses whether explanations are static or interactive, corresponding to presentation format \cite{arya_one_2019}. Reasoning encompasses approaches like abductive and counterfactual reasoning, linking to the goal and content dimensions of explanations \cite{barredo_arrieta_explainable_2020}.

Beyond these aspects, \cite{bellucci_towards_2021} also identified system-specific terminologies, classifying them into interpretable and trustworthy systems. Interpretable systems include characteristics such as black-box-ness, complexity \cite{guidotti_survey_2019}, simulatability, and decomposability. Trustworthy systems emphasize properties such as reliability, fairness \cite{corbett-davies_measure_2023, chen2025decoy, liu2023toward}, and transparency \cite{beaudouin_flexible_2020, ge2024survey}. These characteristics represent the effectiveness of XIS from multiple aspects.

\subsection{The design of explainability}

The design of explainability has received growing interest in XAI research, as it fundamentally shapes how to design XAI so that it can better serve the needs of real users\cite{nguyen_how_2024}. An immediate goal of such work is to comprehend \textit{HOW} to provide explanations effectively for particular user groups, which have been explored in various facets, including the format, structure, and interactivity of the explanations.

Regarding the format of explanations, \cite{mohseni_multidisciplinary_2021} highlighted that explanations can be designed in diverse formats tailored to different user groups, and proposed a classification of formats: visual, verbal, and analytic explanations. Visual explanations employ visual elements to illustrate the model’s reasoning processes, such as visual saliency maps. Verbal explanations deliver information through words, phrases, or natural language, implemented in recommendation systems and robotics. Analytic explanations, on the other hand, present information via numerical metrics and data visualizations to convey model performance or reasoning in a more technical manner. Similarly, when examining the scientific work proposing human-centered approaches to evaluate methods for explainability, \cite{vilone_notions_2021} emphasized the output format of the explanation, including textual, visual, numerical, and mixed. Furthermore, studies such as \cite{weitz2021let} introduced the voice output, exploring how the different modalities of a virtual agent influence the perceived trustworthiness of an AI system. This line of research highlights the growing interest in multi-modal explanation designs, where combining different formats, such as text, visuals, and voice, can enhance user experience and support diverse user needs across application contexts.

The structure of the information from the explainable interface and the interaction design have emerged as crucial aspects of explainability research, as they directly affect how users interact with AI systems. Previous studies have explored how explanatory information can be organized both theoretically and practically. For instance, \cite{nguyen_how_2024} analyzed the design requirements and design outcomes applied by XAI researchers in explainable interfaces through a systematic literature review. They specified design requirements through the concept of visual hierarchy, referring to “the organization of the design elements on the page so that the eye is guided to consume each design element in the order of intended importance,” including global overview, hierarchy overview, multiple instances, focal points, and grouping overview. Regarding how the information structure from the system interface is implemented in the final design, they identified the following information architectures: sequential, matrix, structure, hierarchical, and found that the sequential information architecture was used the most commonly in final designs.

Interaction design represents another crucial aspect of designing for explainability, including whether the explanation user interfaces are interactive and what specific forms that interactive techniques take. Static explanations remain predominant \cite{nguyen_how_2024,bertrand_selective_2023}. However, recent studies found that effective and engaging design is a critical component of any digital product \cite{rogers2023interaction,miller2019explanation} and that there is a growing effort to explore interactive explanation designs \cite{spinner2019explainer,chromik2021human,bertrand_selective_2023}. For example, \cite{nguyen_how_2024} adopted the framework in \cite{rogers2023interaction} to summarize four types of interaction: instructing, conversing, manipulating, and exploring. \cite{bertrand_selective_2023} identified a series of interactive functions that support users to select, mutate, and dialogue with XAI, including clarify, arrange, filter/focus, reconfigure, simulate, compare, progress, answer, and ask. They found that clarify and simulate are the most used interaction techniques, combined with compare, filter/focus and arrange. 

Some studies have also investigated the interplay between explanation formats and interaction designs, identifying patterns in how particular formats tend to be paired with specific interaction types. One of the examples is \cite{arya2019one}’s work, which characterized five concepts of interaction in XAI, including interaction as information transmission, as dialogue, as control, as experience, as optimal behavior, as tool use, and as embodied action. The authors further presented how various modalities (e.g., visual, verbal, chat-based) are embedded in the above interactions. In parallel, \cite{bertrand_selective_2023} observed that charts and textual explanations are the most commonly used elements in XAI design, with tables proving particularly effective for supporting \textit{Filter/Focus} and \textit{Compare} interactions, and textual explanations frequently accompanied by \textit{Clarify} interactions that allow users to control the information displayed on demand.

\subsection{Measurement of explainability}
\subsubsection{Measurement dimension}
Measurement dimensions of explainability can be grouped into two categories: explanation-grounded dimensions, which assess the properties and quality of the explanations themselves, and effects-grounded dimensions, which evaluate the impacts of explainability on users and system-related outcomes. For instance, \cite{haque_explainable_2023} adopted the information quality dimensions to conceptualize the explanation quality dimensions, including format, completeness, accuracy, and currency. They also categorized explainable AI effects into trust, transparency, usability, understandability, and fairness. \cite{hoffman_metrics_2019} highlighted four major classes of measures corresponding to the XAI explaining process: the goodness of explanations, explanation satisfaction, mental models, and performance. The first two are related to the explanation itself, while the latter two are assessed for outcomes. Similarly, \cite{lofstrom_meta_2022} identified that evaluating the use of explanation methods comprises three aspects: model, explanation, and user. The user aspect includes four criteria: trust, appropriate trust, bias detection, and explanation satisfaction. The explanation aspect encompasses five criteria: fidelity, identity, separability, novelty, and representativeness. Regarding the model aspect, there are three evaluation criteria: performance, fairness, and reliability. 
 
 Some other studies take a broad perspective by compiling evaluation measures derived from prior work. For instance, \cite{mohseni_multidisciplinary_2021} identified broad categories of evaluation measures for XAI systems and algorithms, including user mental model, explanation usefulness and satisfaction, user trust and reliance, human-AI task performance, and computational measures. \cite{nguyen_how_2024} conducted a focused review of how explainability interfaces are evaluated and found that prior studies have used multiple metrics, including understandability, task performance (e.g., task accomplishment), transparency, effectiveness, efficiency, user trust, and satisfaction. Additionally, \cite{nunes_systematic_2017} aimed to report on the evaluation of explanations in past user studies, similar to the present review. They presented the question topics in subjective perception questionnaires using a tag cloud, in which transparency, usefulness, usability, satisfaction, and trust are the main aspects evaluated in user studies.
 
\subsubsection{Measurement approach}
Since the approaches are developed to address different evaluation goals of explainability, various methods have been proposed to assess both the quality of explanations and their effects based on these goals. Recent studies have widely recognized a dual categorization for evaluating methods of explainability \cite{vilone_notions_2021,rong_towards_2023,pietila_when_2024,bibal2016interpretability,doshi2018considerations}. This categorization features a different naming scheme that distinguishes between objective evaluations and human-centered evaluations. For instance, \cite{vilone_notions_2021} proposed that there are two main ways to evaluate explainability methods in XAI research: (1) objective evaluations, i.e., objective metrics and automated approaches to evaluate methods for explainability. A thorough review of metrics that can be computed without human subjects is available in \cite{nauta2023anecdotal}; (2) human-centered evaluations, employing a human-in-the-loop approach that relies on human participants to collect feedback and evaluations.  \cite{pietila_when_2024} reviewed the current metrics of XAI systems in the healthcare domain, suggesting two classes of metrics: metrics without a human-in-the-loop and metrics with a human-in-the-loop. Similarly, reviews by \cite{rong_towards_2023} divided the evaluation measures into two groups: human-grounded measures and functionally-grounded metrics.  

Regarding human-centered evaluations, which focus on user studies concerning XAI, \cite{vilone_notions_2021} identified two categories based on the nature of the questions posed to participants. Qualitative studies typically collect detailed, in-depth feedback from end-users, providing valuable insights into how participants engage with reasoning processes and their subjective experiences when interacting with explanations. Common methods in this category include think-aloud protocols, semi-structured interviews, free-text analyses, etc \cite{bertrand_selective_2023, hoffman_metrics_2019}. In contrast, quantitative studies typically employ methods such as experiments and surveys to generate data that can be quantitatively measured and statistically analyzed. Commonly used measurement tools include various scales (e.g., System Causability Scale, NASA-TLX, self-developed Likert scales), subjective ratings, and objective measures \cite{vilone_notions_2021, pietila_when_2024}. The indicators measured often included human performance and interaction patterns, such as task completion time, accuracy, and interaction time \cite{bertrand_selective_2023,rong_towards_2023,saeed_explainable_2023}. Additionally, a considerable number of studies utilize mixed-methods designs, combining qualitative and quantitative data to provide a holistic view of XAI contexts.

\subsection{Summary}

In summary, the existing research largely addresses the question of explainability in three key areas. First, the concept of explainability varies across the research community, with related terms such as interpretability, understandability, and transparency frequently used interchangeably. These conceptualizations reflect distinct disciplinary priorities and methodological perspectives, leaving a gap for a unified, operational definition. Second, concerning the design of explainability, prior studies have explored diverse explanation formats, interaction mechanisms, and presentation strategies. Earlier research has examined how visual, verbal, and analytic explanations are implemented, as well as how interaction techniques shape users’ understanding and trust in AI systems. These designs are frequently customized for various user types, needs, and environments. However, no existing review has systematically mapped how these design elements correspond to specific interface types and how they are adapted to accommodate the expectations of different audiences within interactive information systems. Third, we reviewed the related work on dimensions and approaches for measuring explainability. Prior work has proposed a dual categorization that distinguishes between objective evaluations based on computational metrics and human-centered evaluations involving user studies to assess explanations and their effects on users. No review to date has systematically considered how to measure explainability itself in user studies of interactive information systems. Moreover, beyond listing measures, there remains a lack of higher-level categorization that organizes these diverse metrics into meaningful groups, limiting our understanding of identifying consistent evaluation practices across studies.  

Against this background, this survey paper addresses these gaps by offering a comprehensive synthesis of \textit{human-centered} explainability research in interactive information systems. Specifically, it examines how explainability has been conceptualized, how explanation features have been designed and integrated into system interfaces and user interactions, and how explainability has been measured and evaluated in empirical user studies. This review seeks to provide a clearer and more structured understanding of human-centered explainability, while also guiding future design, implementation, and evaluation practices in this rapidly evolving research area.

\section{Methods}\label{methods}
\begin{figure}
    \centering
    \includegraphics[width=0.9\linewidth]{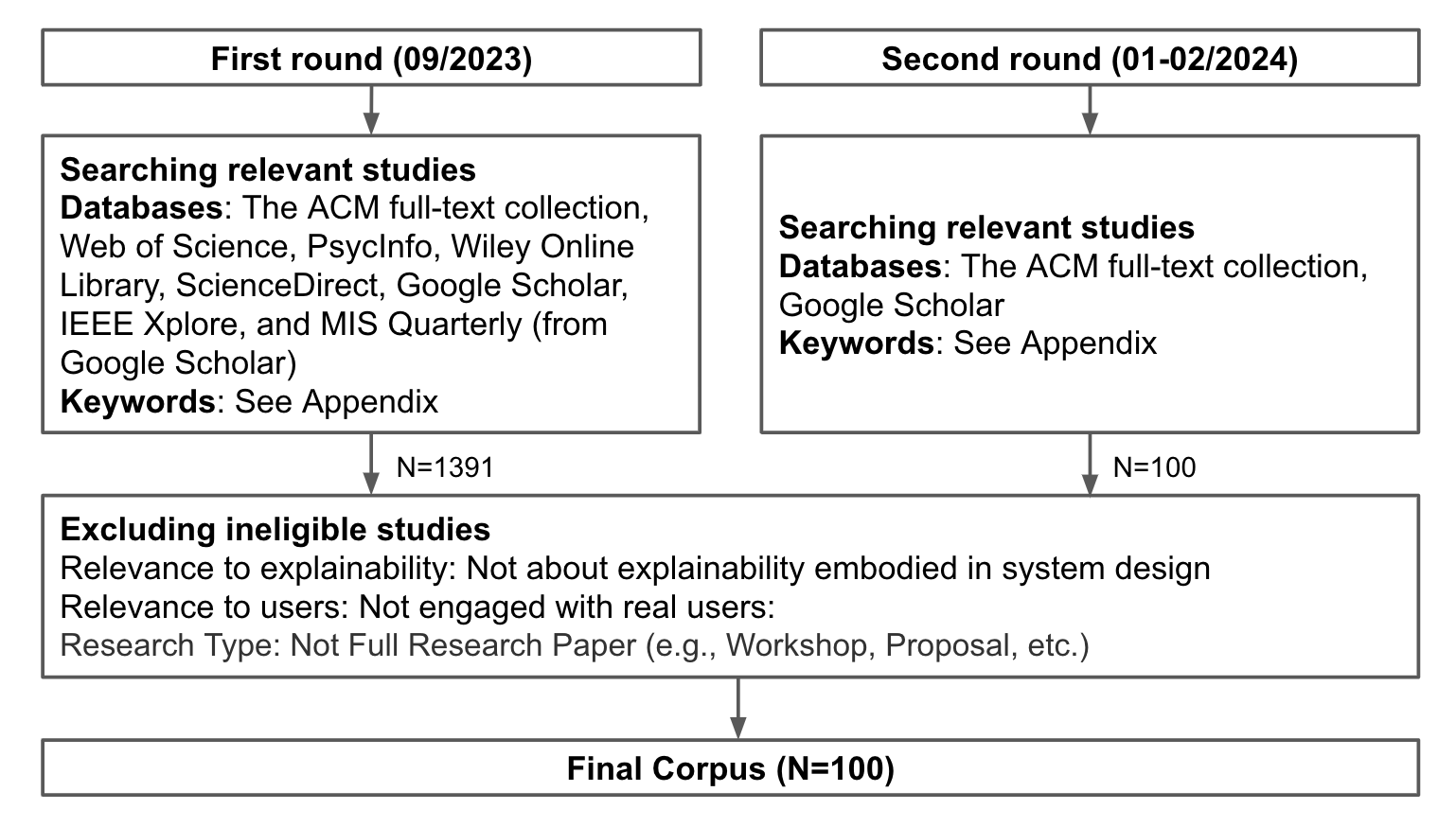}
    \caption{Number of included articles}
    \label{fig:prisma}
\end{figure}

We chose a systematic review method to gain a thorough understanding of human-centered explainability. We adhered to the guidelines set forth by the Preferred Reporting Items for Systematic Review and Meta-Analysis (PRISMA) \cite{page_prisma_2021}. To outline a robust PRISMA process, in this section, we will present the literature review search (Section \ref{subsec:literature search}), review inclusion and exclusion (Section \ref{subsec:criteria}), study identification (Section \ref{subsec:study identification}), and the qualitative coding methodology (Section \ref{subsec:data extraction}). Figure \ref{fig:prisma} demonstrates the whole literature searching and screening process.

\subsection{Literature Search} \label{subsec:literature search}
We conducted our search across eight scholarly databases: the ACM Full-Text Collection, Web of Science, PsycInfo, Wiley Online Library, ScienceDirect, Google Scholar, IEEE Xplore, and MIS Quarterly (via Google Scholar). These databases were chosen as they encompass scholarly publications from academic disciplines such as human-computer interaction and information science, which are most relevant to research on \textit{Human-Centered Explainability}. The search queries-\textit{AI OR search OR information retrieval OR recommendation}-combined with the following keywords and their variations: user, explainability, and interactive information system. Specific queries were tailored to each database and are detailed in Table \ref{query} of the Appendix.

Searches targeted titles, keywords, and abstracts in all databases except Google Scholar. For Google Scholar, the first and second authors utilized the "with all of the words" field to apply the search queries. We initially screened the top 20 results. If relevant articles not covered in other databases were identified, the screening was expanded to the top 50 and subsequently the top 100 results. No starting year was set for the search, ensuring a comprehensive coverage of the literature.

One author conducted the searches, removing duplicates by examining titles, abstracts, and DOI numbers. The initial batch search took place in September 2023. To collect missing literature with different keywords or related to the newly emerged AI topic, we conducted the second search on ACM and Google Scholar in January 2024. This search included keywords such as ``interpretability,'' ``scrutability,'' (to replace ``explainability'' in the original query) and ``Generative AI (GenAI)'' (to replace``interactive information system``). We collected 1391 papers in the first search and 100 papers in the second search.

\subsection{Inclusion and Exclusion Criteria} \label{subsec:criteria}
\begin{figure}
    \centering
    \includegraphics[width=0.7\linewidth]{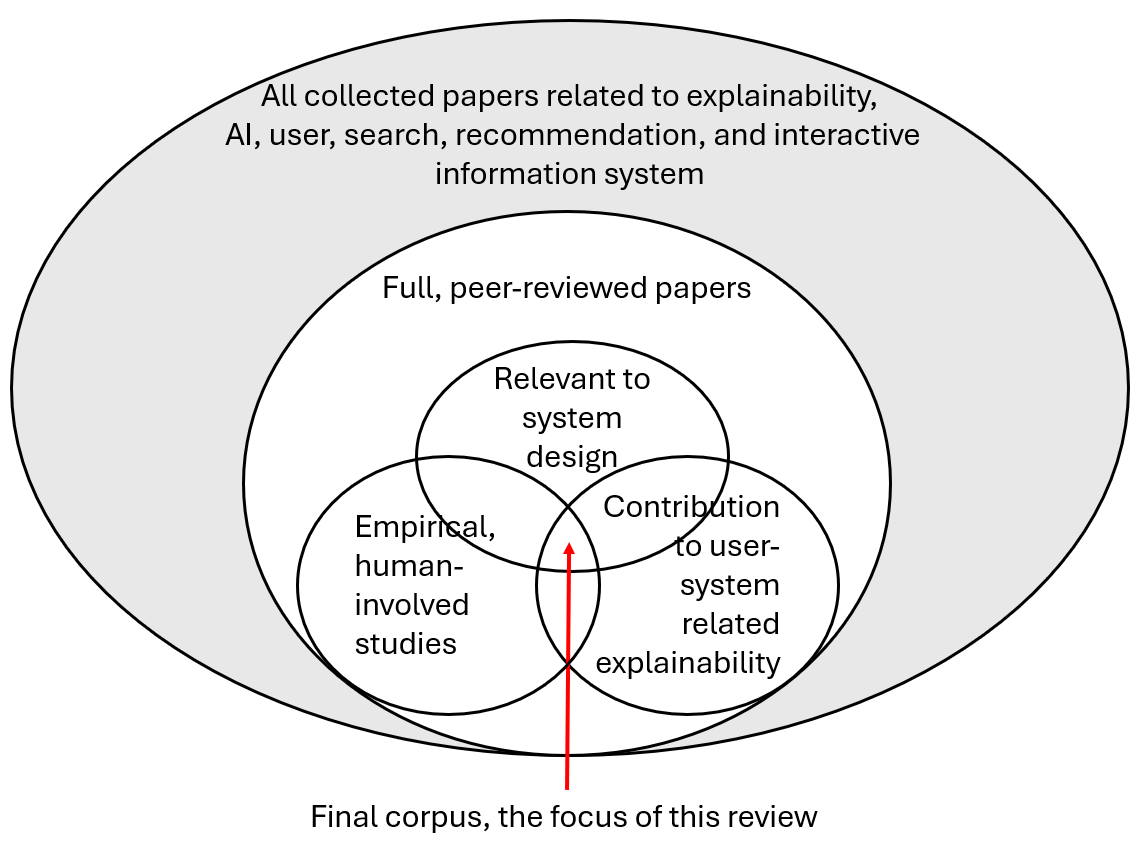}
    \caption{Literature search: Inclusion and exclusion criteria.}
    \label{fig:boundary}
\end{figure}

As shown in Figure \ref{fig:boundary}, we screened the literature based on the following \textit{inclusion and exclusion criteria}: 
1) Empirical studies that involve real users in system design and evaluation. Papers that only include data-driven simulations without human evaluation were excluded. For example, \cite{jeyasothy2023general} proposed a method to integrate user knowledge into explanation to make it understandable. However, it mainly utilized existing dataset without further user studies or engagement, so it was excluded. \cite{setchi2020explainable} shared perspectives in explainable human-robot interactions without empirical studies, so it was excluded.
2) The findings can contribute to the understanding of user needs, perception, and evaluation of explainability. For example, \cite{he2021explainable} investigated explainability in UAV algorithms. Although it involves human user evaluations, it is not related to user needs or perceptions, so it was excluded.
3) The findings shed light on the way that explainability is embodied in system design. For example, \cite{el2023sand} mainly focused on the self-explainable interactive method, which is not related to the information systems, so it was excluded.
4) Full papers written in English that were published in peer-reviewed journals or conference proceedings, as a majority of recent works were published in English venues. Non-full or non-peer-reviewed papers, such as workshop papers and extended abstracts, were excluded.

\subsection{Study identification} \label{subsec:study identification}
Using the inclusion and exclusion criteria, we filtered and analyzed papers. Three researchers separately coded one-third of articles and cross-checked with each other. Disagreements were addressed through discussions. In total, we included 100 articles. Among them, 17 included more than one user study, resulting in a total of 121 reported user studies.

\subsection{Data extraction}\label{subsec:data extraction}
The data extraction process was conducted using Google Sheets. Key study characteristics were manually recorded, including publication venues, authors' geographical locations based on their affiliations, and participant demographics, including  the number of participants, age, gender, education, race, income, and occupation. The first three authors each performed the initial data extraction on one third of the papers and then cross-checked the coding results with each other. Discrepancies were addressed through group discussion. We also referred to previous studies to develop the codebook for research method \cite{creswell_mixed_2019}, participants' level of knowledge on AI \cite{bertrand_selective_2023} or specific domain \cite{bertrand_selective_2023}, and the domain to which explainable interface is applied \cite{lai_towards_2021, a_systematic_2023}. Table \ref{tab:codebook_study_feature} presents the definition of each code.

\begin{table}
\caption{The codebook for study characteristics}
\label{tab:codebook_study_feature}
\setlength{\arrayrulewidth}{0mm}
\centering
\small
\begin{tabular}{|p{3cm}|p{10cm}|}
\toprule
    \multicolumn{2}{|p{13cm}|}{\textbf{Research method}}\\
        Qualitative & The research method that gather and analyze non-numerical data, e.g., interview and focus groups.\\
        Quantitative & The research method that gather and analyze numerical data, e.g., experiment and survey. \\
        Mixed method & The research design that combines quantitative and qualitative research methods to draw on the strengths of each. \\
    \multicolumn{2}{|p{13cm}|}{\textbf{Participants' level of knowledge on AI/domain}}\\
        Expert & Participants are qualified as expert if they have sufficient knowledge on AI or related domains (e.g., healthcare professionals and lawyers). \\
        Novice & Participants have limited knowledge of AI or related domains. \\
        Uninitiated & Participants have no knowledge of AI or related domains. \\
        Unreported & The paper didn't reported participants' level of knowledge on AI or related domains. \\
    \multicolumn{2}{|p{13cm}|}{\textbf{Application domain and examples}}\\
        Academia & The explainability design targeted academia scholars, e.g., scholar recommendation explanation \cite{tsai_evaluating_2019}. \\
        Architecture & The explanation that helps user understand specific architecture design recommendation \cite{cox_explainable_2018}. \\
        Education & The explanation that help students or teachers adopt systems, e.g., intelligent tutoring system \cite{conati_toward_2021}, to assist their learning. \\
        Engineering & The explainability design concerned with the design, building, and use of engines, machines, and structures, such as energy efficiency recommendation \cite{sardianos_emergence_2021} and air-handling unit faults diagnosis \cite{meas_explainability_2022}. \\
        Food safety & The explainability design supporting the identification of safe food, e.g., edible mushroom identification \cite{leichtmann_effects_2023}.\\
        Finance \& Business & The explanation that helps user to make financial and business decisions, such as hotel price prediction \cite{westphal_decision_2023} and income prediction\cite{tesic_can_2022}.\\
        Generic & The explanability design that can be applied to different domains, such as text classification \cite{riveiro_thats_2021}, pfishing attack \cite{desolda_explanations_2023}, programming \cite{wijekoon_user-centred_2023}, robotics \cite{cruz_evaluating_2022}, facial image recognition \cite{john_explaining_2023}, and unmanned aerial vehicle \cite{mualla_quest_2022}. \\
        Law \& Civic & The explanation aims to assist legal and civic activities, such as criminal investigation \cite{hepenstal_developing_2021} and legal case reasoning \cite{collenette_explainable_2023}. \\
        Leisure & The explanation that helps with users' entertainment, such as the explanations for book recommendation \cite{dey_evaluating_2021} and movie recommendation \cite{lully_enhancing_2018}.\\
        Medical \& Health & The explanation that help with specific tasks of patients, healthcare providers, and other stakeholders in medical or general health contexts, such as the explanation of COVID information \cite{maruf_influence_2023}, COVID diagnosis \cite{goel_effect_2022}, and personal lifestyle recommendation \cite{dragoni_explainable_2020}.\\
        Transportation & The explainability design that help users to understand and use systems for transportations, such as intelligent vehicles \cite{graefe_human_2022}.\\
    \bottomrule
\end{tabular}
\end{table}

We applied different analysis approaches to answer each research question. For the definition of explainability (RQ1), we extracted the original definition of explainability-related concepts from the included papers into Google Sheet, where we identified three groups of definitions regarding \textit{explanation}, \textit{explainability}, and \textit{XAI}. We then conducted textual analysis using Python to shed light on the similarity and differences among the three concepts (see Section \ref{subsec:exp_def}). 

We further categorized the keywords based on different facets of the definition, focusing on aspects including \textbf{where} (the context in which an explanation is needed), \textbf{why} (the goal that the explanation aims to achieve), \textbf{whom} (the target audience that the explanation is intended for), \textbf{what} (the content included in the explanation), and \textbf{how} (the actions or effects that the explanation has on the user). These dimensions help illuminate various perspectives, such as different roles and functions within the concept of explainability, providing a comprehensive framework for analysis.

For the design elements of explainability (RQ2), an codebook (see Table \ref{tab:category_design_elements}) was developed referring to existing design catalogs \cite{google_components_nodate,ribecca_data_nodate}; we took screenshots of all interfaces all these articles, one researcher coded the screenshots following the codebook and another researcher validated the coding results, discrepancies were addressed through group discussion. We then conducted descriptive analysis of the coding results (see Section \ref{subsec:exp_interface}). 
\begin{table}[h]

   \caption{Categories of explainability design}
    \label{tab:category_design_elements}
    \centering
    \small
    \begin{tabular}{p{0.12\textwidth}p{0.18\textwidth}p{0.58\textwidth}}
        \toprule
        \textbf{Category} & \textbf{Sub-category} & \textbf{Definitions} \\ 
        \midrule
        
    Interactivity & Non-interactive &  A pre-generated prototype or explanations without interactive functions. e.g., a screenshot of an explanation interface \cite{westphal_decision_2023}. \\
    &Interactive  & A prototype with interactive functions \cite{meske_design_2023, graefe_human_2022}.\\
    
    Modality & Text-based & The explanation is conveyed through text, e.g., a textual explanation \cite{tesic_can_2022}.\\
    &Graphical-based & The explanation is conveyed through visual representations, e.g., the data visualization such as bar chart, word cloud, and radar graph \cite{tsai_evaluating_2019}.\\
    &Video-based & The explanation is presented through video media, such as a video loop \cite{singla_explaining_2023}.\\
    &Voice-based & The explanation is conveyed through the voice, e.g., natural language in voice from wearable glasses \cite{danry_wearable_2020}.\\
    
    Interface type & Chatbot & A software program designed to simulate conversation in natural language with human users \cite{dragoni_explainable_2020, musto_linked_2019}.\\
    &Dashboard & A type of graphical user interface (GUI) which often provides at-a-glance views of data relevant to a particular objective or process \cite{meske_design_2023, tandon_surfacing_2023}.\\
    &  Desktop app & A software program that runs on a desktop computer or a laptop \cite{nourani_investigating_2020, subramonyam_protoai_2021}.\\
    &  Game & A game played using a computer, typically a video game \cite{sreedharan_using_2021}.\\
    & Wearable glass & Eyeglasses that have built-in technology that adds digital information to the wearer's view \cite{danry_wearable_2020}.\\
    & Mobile app & A software program that runs on a mobile device, like a smartphone or tablet \cite{jiang_who_2022, kim_stakeholder_2024}.\\
    & Pop-up message & A brief message that appears on a website or mobile app to inform or prompt users to take an action \cite{desolda_explanations_2023, graefe_human_2022}.\\
    &  Virtual reality & A computer-generated environment that simulates reality \cite{manger_providing_2023}.\\
    &  Web-based tool & A software application that runs in a web browser and can be accessed over the internet \cite{dey_evaluating_2021, collenette_explainable_2023}.\\
    \bottomrule
    \end{tabular}
\end{table}

For the measure of explainability (RQ3), we first extracted the constructs measured and corresponding items into Google Sheet, we then categorized these constructs and items referring to a past data quality framework \cite{wang_beyond_1996} and XAI evaluation measures \cite{mohseni_multidisciplinary_2021,vilone_notions_2021}. The previous framework and measures suggest multidimensional, user-centered approaches for explainability studies. Following these approaches, we extracted categories for explainability measures, which include intrinsic, format and presentation, usability, experiential, ethics, and interaction with the explanation. The first five are subjective measurements, while the last focuses on objective metrics. The codebook is shown in Table \ref{tab:measurement definition} (see Section \ref{subsec:exp_measure} for corresponding results)

\begin{table}[h]
   \caption{Categories of measured constructs}
   \label{tab:measurement definition}
    \centering
    \small
    \begin{tabular}{p{0.35\textwidth}p{0.6\textwidth}}
        \toprule 
        \textbf{Category} & \textbf{Definitions} \\ 
        \midrule
        Intrinsic  & The inherent attributes of explanations in its own right.\\ 
        Format and presentation  & The way that how information is structured, styled, and delivered to users.\\
        Usability  & Refers to the ease with which users can understand, interact with, and derive value from the explanation provided.\\
        Experiential & Focuses on the subjective user experience and how users perceive and interact with the explanations.\\
        Ethics  & Whether the ethical considerations were given attention.\\
        \RaggedRight Interaction with the Explanation & Objective metrics to measure if and how the users engaged with the explanations. \\
        \bottomrule
    \end{tabular}
\end{table}

\section{Findings}
To report the findings of our systematic review, we begin by providing an overview of the descriptive statistics for the 121 studies included in this review (Section \ref{subsec:study_characteristics}). Following this, we present a synthesis of our coding themes to address each research question (Section \ref{subsec:exp_def} - \ref{subsec:exp_measure}). 

\subsection{Study characteristics} \label{subsec:study_characteristics}
\subsubsection{Participant characteristics}
We provided an overview of participant characteristics in 121 studies reported in the 100 included articles. The sample size of participants ranged from 3 to 680 (\textit{M} = 95.37, \textit{SD} = 126.10). Most of these studies (\textit{N} = 81) involved fewer than 100 participants. We described participant characteristics as follows:

\textbf{Age. }
Across the 121 studies from 100 included articles, there were significant variations in the age ranges of the participants.
Specifically, 42 studies reported the age range of participants, the minimum age was between 16 and 25 (\textit{M} = 19.98, \textit{SD} = 2.40) and the maximum age was between 23 and 84 (\textit{M} = 50.43, \textit{SD} = 14.81). 28 studies reported the mean age of participants, which is between 20.7 and 61.0 (\textit{M} = 34.71, \textit{SD} = 9.05); 61 studies didn't report participants' age. 

\textbf{Gender. }
Of the 121 studies, 56 reported participants' gender. The average percentage of female participants is 47.56\% (\textit{SD} = 17.37\%). 46.43\% studies (\textit{N} = 26) had a sample with more than 50\% being women participants. 

\textbf{Level of knowledge on AI. } 
Among the 121 studies, 34 reported participants’ knowledge levels of AI. Of these, 22 studies specified participants as having no AI knowledge (\textit{N} = 4), basic knowledge as novices (\textit{N} = 12), or sufficient knowledge as experts (\textit{N} = 27); the rest of them (\textit{N} = 12) recruited participants with various levels of knowledge on AI.

\textbf{Level of domain knowledge. }
57 studies reported participants' knowledge on the domain that the system targeted to, including domain experts (\textit{N} = 35), with various level of domain knowledge (\textit{N} = 18), and with no domain knowledge (\textit{N} = 2).

\subsubsection{Research method}
Among these 121 included studies, more than half of them (\textit{N} = 65) adopted the quantitative approach; 20 of them used the qualitative approach; and 36 of them applied mixed-method approaches. 

\textbf{Quantitative approach.} Experiment (\textit{N} = 46) and survey (\textit{N} = 19) were two major design in these approach. For experiment studies, 22 of them were conducted on the cloud-sourcing platform (e.g., Amazon Mechanical Turk \cite{wijekoon_user-centred_2023} and Prolific \cite{riveiro_thats_2021}), six of them conducted the experiments through online surveys (e.g., \cite{pumplun_bringing_2023} distributed survey through their direct contacts), one conducted field experiment (\cite{dragoni_explainable_2020} invited users to use their systems for four days in order to obtain their evaluation in real life). For survey studies, one of them combined survey with eye-tracking to enrich the data collection \cite{dey_evaluating_2021}. 

\textbf{Qualitative approach.} This group of studies mainly applied interview (\textit{N} = 19) to collect data. The rest design included workshop \cite{sun_investigating_2022} and open-ended question survey \cite{cox_explainable_2018}. 

\textbf{Mixed method approach.} We identified the following combinations of methods in this group of studies: (1) Experiment combined with methods to collect qualitative feedback (\textit{N} = 17), including open-ended questions in surveys (\textit{N} = 11, e.g., \cite{riveiro_thats_2021, mukhtar_explaining_2023}, asking general comments (\textit{N} = 5, e.g., \cite{schultze_explaining_2023}) or textual feedback (\textit{N} = 1, \cite{draws_explainable_2023}), written reflection (\textit{N} = 1, \cite{olson_counterfactual_2021}), think-aloud (\textit{N} = 1, \cite{naiseh_how_2023}), and interviews (\textit{N} = 1, \cite{manger_providing_2023}). (2) Survey combined with open-ended question (\textit{N} = 11, e.g., \cite{meske_design_2023, novak_transferring_2022}), interviews (\textit{N} = 3, e.g., \cite{butz_investigating_2022}), and general comments (\textit{N} = 3, e.g., \cite{anjomshoae_context-based_2021}). Additionally, one study applied a combination of card-sorting and interviews to understand participants' ranking of different visualizations and corresponding qualitative feedback \cite{tsai_evaluating_2019}. 

\subsection{Explainability Definition} \label{subsec:exp_def}

As the papers used different terms in their definition, we examined how variations in keywords for explanation, XAI, and explainability contribute to differences among the definitions of these three definition terms. Table \ref{tab:definition} lists the keywords from definitions of the three terms. In addition, we separated the keywords into five categories: \textbf{where}, \textbf{why}, \textbf{whom}, \textbf{what}, and \textbf{how}. 

\begin{table}
\centering
\small
\caption{Definitions Keywords in three definition terms. (\textit{The full list of word frequency is in Appendix Table \ref{ap-tab:definition}})}
\label{tab:definition}
\begin{tabular}{lr|lr|lr} \hline
\multicolumn{1}{c}{\textbf{XAI}} & \multicolumn{1}{c}{} & \multicolumn{1}{c}{\textbf{Explanation}} & \multicolumn{1}{c}{} & \multicolumn{1}{c}{\textbf{Explainability}} & \multicolumn{1}{c}{} \\ \hline
\multicolumn{1}{c}{\textit{Where}} & \multicolumn{1}{c|}{} & \multicolumn{1}{c}{\textit{Where}} & \multicolumn{1}{c|}{} & \multicolumn{1}{c}{\textit{Where}} & \multicolumn{1}{c}{} \\
AI & 45 & \textbf{decision} & 10 & \textbf{model} & 9 \\
\textbf{system} & 21 & \textbf{model} & 7 & \textbf{system} & 9 \\
\textbf{decision} & 13 & \textbf{system} & 4 & \textbf{decision} & 6 \\
\textbf{model} & 13 & \multicolumn{1}{c}{\textit{Why}} & \multicolumn{1}{l|}{} & AI & 5 \\
ML & 4 & \textbf{understand} & 7 & ML & 3 \\
blackbox & 4 & help & 5 & \multicolumn{1}{c}{\textit{Why}} & \multicolumn{1}{l}{} \\
\multicolumn{1}{c}{\textit{Why}} & \multicolumn{1}{l|}{} & \multicolumn{1}{c}{\textit{Whom}} & \multicolumn{1}{l|}{} & \textbf{understand} & 9 \\
\textbf{understand} & 15 & \textbf{user} & 7 & ability & 7 \\
transparency & 8 & human & 3 & interpretable & 3 \\
interpretable & 5 & \multicolumn{1}{c}{\textit{What}} & \multicolumn{1}{l|}{} & clearer & 2 \\
trust & 4 & recommendation & 10 & comprehensible & 2 \\
help & 3 & reason & 6 & \multicolumn{1}{c}{\textit{Whom}} & \multicolumn{1}{l}{} \\
accuracy & 2 & feature & 5 & human & 9 \\
improve & 2 & item & 4 & \textbf{user} & 8 \\
\multicolumn{1}{c}{\textit{Whom}} & \multicolumn{1}{l|}{} & information & 3 & \multicolumn{1}{c}{\textit{What}} & \multicolumn{1}{l}{} \\
\textbf{user} & 13 & insight & 3 & information & 7 \\
human & 8 & justification & 3 & item & 4 \\
\multicolumn{1}{c}{\textit{What}} & \multicolumn{1}{l|}{} & alternative & 2 & recommendation & 4 \\
recommendation & 6 & behavior & 2 & xxx-based & 3 \\
specific & 4 & causality & 2 & output & 3 \\
prediction & 3 & classification & 2 & prediction & 3 \\
result & 3 & quality & 2 & specific & 3 \\
technology & 3 & specific & 2 & everything & 2 \\
inner workings & 3 & \multicolumn{1}{c}{\textit{How}} & \multicolumn{1}{l|}{} & feature & 2 \\
output & 2 & describe & 3 & internal & 2 \\
reason & 2 & present & 3 & reason & 2 \\
\multicolumn{1}{c}{\textit{How}} & \multicolumn{1}{l|}{} &  & \multicolumn{1}{l|}{} & \multicolumn{1}{c}{\textit{How}} & \multicolumn{1}{l}{} \\
Reveal & 2 &  & \multicolumn{1}{l|}{} & present & 2 \\ \hline
\multicolumn{3}{l}{\scriptsize The main keywords shared by three terms are in bold face.}
\end{tabular}
\end{table}

The results show that certain high-frequency keywords overlap across all three definition terms, including ``system'', ``decision'', ``model'', ``understand'', and ``user''. These shared keywords point to a fundamental focus on clarifying how explanations/explainability relates to systems, users, and decision-making, which is the main focus of the definition of explainability in the three terms. These words indicate an emphasis on the context in the system and its decisions and models where explanations are needed, the underlying goal of fostering understanding, and the user who receives the explanations.

\begin{figure}
    \centering
    \includegraphics[width=0.85\linewidth]{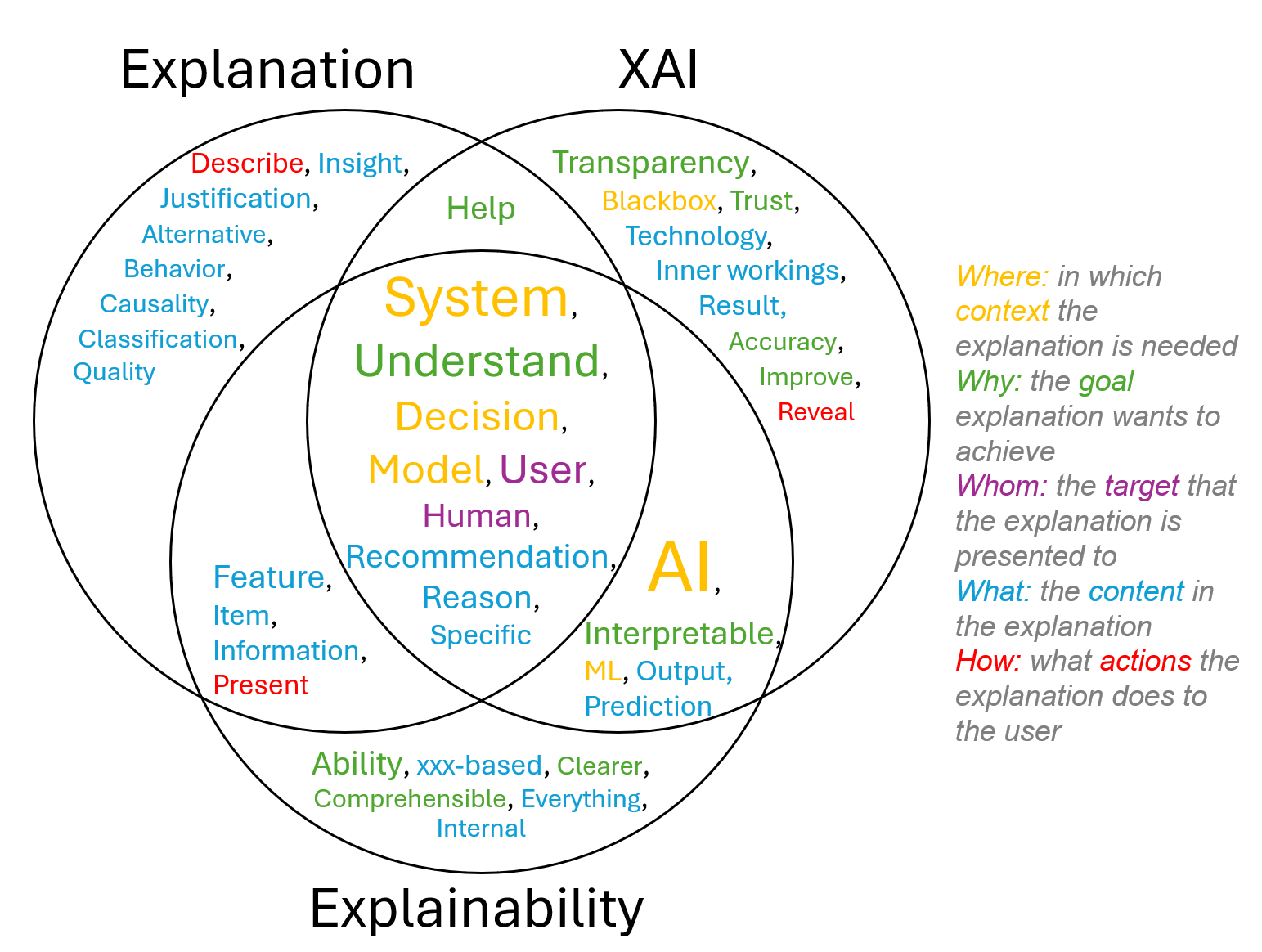}
    \caption{Definition keyword distribution in three definition terms. The colored texts indicate different types of the keyword. The font size of the keyword is adjusted according to its frequency.}
    \label{fig:definition}
\end{figure}

Despite these commonalities, each definition term exhibits distinctive emphases. Figure \ref{fig:definition} presents the Venn diagram of the definition keywords. The \textit{Explanation} term tends to focus on rich, content-driven definitions (\textbf{what}), highlighting elements such as ``justification'', ``alternatives'', ``behaviors'', ``causality'', and ``classification''. By contrast, the \textit{XAI} term widens the scope to include comprehensive aspects (\textbf{where}, \textbf{why}, and \textbf{what}) related to ``transparency'', and ``trust'', as well as explicitly referencing ``black-box'' models and the ``inner workings''. Except for the major commonalities, the overlap between only explanation and XAI is limited with the keyword ``help'', suggesting divergent priorities. \textit{Explanation} remains content-centric, whereas \textit{XAI} incorporates a broader set of objectives related to the goal of the explanability.

The \textit{Explainability} term appears to bridge these two terms above. It overlaps with \textit{Explanation} in terms of content-related keywords (\textbf{what}), such as ``features'', ``items'', and ``information'', indicating a shared focus on the content-oriented definition of providing meaningful, detailed explanations. It also overlaps with \textit{XAI} in terms of context-oriented and goal-related keywords (\textbf{where} and \textbf{why}) like ``AI'', ``ML'', and ``interpretable'', reflecting an integrated perspective that addresses not only what information is provided but also how and why it should be made understandable. This bridging role suggests that \textit{Explainability} merges the content-rich foundation of \textit{Explanation} with the broader aims of \textit{XAI}.

Interestingly, the highest frequency keyword ``AI'' exists in the \textit{XAI} and \textit{Explainability} definitions but is absent from \textit{Explanation}. This pattern suggests that the traditional notion of \textit{Explanation} may have a long-standing conceptual heritage in various fields but may not need to be restricted to AI contexts, while The concepts of \textit{XAI} and \textit{Explainability} are more distinctly grounded in the challenges posed by complex AI and ML systems, which is closely tied to the need for explanation. In addition, in terms of \textbf{how} explainability is conveyed, we identified keywords such as ``describe'', ``reveal'', and ``present'' in the definitions. These actions do not refer to the specific form, like visual or textual explanations, but reflect the intended actions that explanations perform on behalf of the system to achieve the goal to the user.

\subsection{Explainability interface} \label{subsec:exp_interface}
Out of the 100 articles included in our analysis, 85 detailed their explanation design. The interface design was analyzed along three dimensions: interactivity, modality, and interface type. 

\textit{Interactivity} refers to whether the explanation interfaces evaluated in user studies allowed users to interact with the explanation content actively. Among the 85 articles, 44 utilized interactive interfaces, which are prototypes enabling participants to adjust, query, or manipulate explanations through interactions with them, such as chatbots and mobile applications. Another 42 studies investigated non-interactive interfaces, which provide static, pre-generated explanations without user control, such as prepared text or prototype screenshots. Notably, one article employed both types of interfaces in multiple user studies \cite{musto_linked_2019}, resulting in a total of 86 explanation setups analyzed across 85 papers.

\begin{figure}
    \centering
    \includegraphics[width=0.9\linewidth]{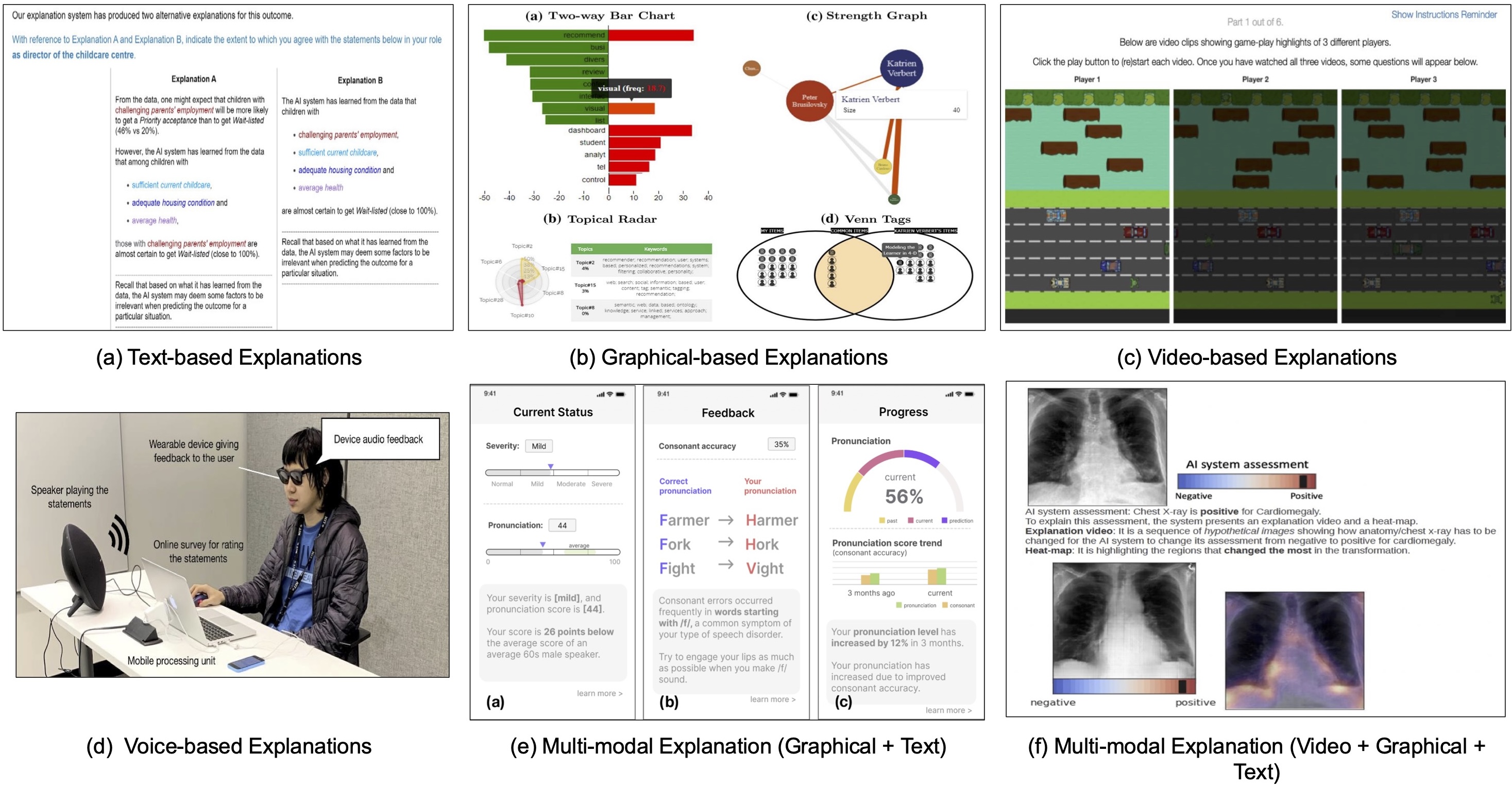}
    \caption{Examples of explainability modalities. Top Row (Left to Right): \cite{maruf_influence_2023,tsai_effects_2021,sequeira_interestingness_2020}. 
    Second Row \cite{danry_wearable_2020,kim_stakeholder_2024,singla_explaining_2023}}
    \label{fig:examples-explanations}
\end{figure}

\textit{Modality} refers to the medium through which the explanation is delivered to the user. We identified four primary modalities in the reviewed studies: text-based, graphical-based, video-based, and voice-based explanations. Additionally, some studies employed multi-modal designs, combining two of these modalities to enhance explanatory effectiveness. Examples of various types of modalities are presented in Figure \ref{fig:examples-explanations}. A considerable number of studies adopted multi-modal designs, most often combining graphics and text to improve explanatory clarity and engagement. The remaining instances utilized a single modality, accompanied by formatting techniques. For example, text-based explanations often emphasize key information and improve readability through visual techniques such as bold, italic, or underlining. Similarly, graphical explanations varied in their visual strategies, including the use of color coding and highlighting to increase explainability. Video-based explanations, although relatively uncommon, appeared in a small subset of studies (N=3) to present animated or dynamic content. These explanations typically illustrated sequential system behaviors and the transformation of decision outcomes (e.g., from negative to positive \cite{singla_explaining_2023}) in a clear and intuitive manner. Only one paper delivered explanations through spoken natural language \cite{danry_wearable_2020}. Despite their potential advantages, particularly in mobile, assistive, or hands-free contexts, auditory modalities remain significantly underexplored within the current explainability research.

\textit{Interface type} refers to the platform or application environment in which explanations are delivered. Nine categories of interface types were identified. Among the 44 studies employed interactive interfaces, the vast majority were graphical user interfaces (GUI) (N=51), with dashboards being the most common (N=21), followed by desktop applications (N=5), chatbots (N=4), web-based tools (N=4), mobile apps (N=3), pop-up messages (N=2), computer programs (N=1), VR applications (N=1), and games (N=1). Only one study employed a voice user interface (VUI), using a wearable smart glasses system to provide explanations via voice-based natural language \cite{danry_wearable_2020}. Two studies reported their interactive functions without specifying the exact interface type. The identified interface types highlight the diverse and evolving contexts in which explainability is embedded, spanning from conventional dashboards to immersive VR experiences and emerging wearable technologies.

\begin{figure}
    \centering
    \includegraphics[width=\linewidth]{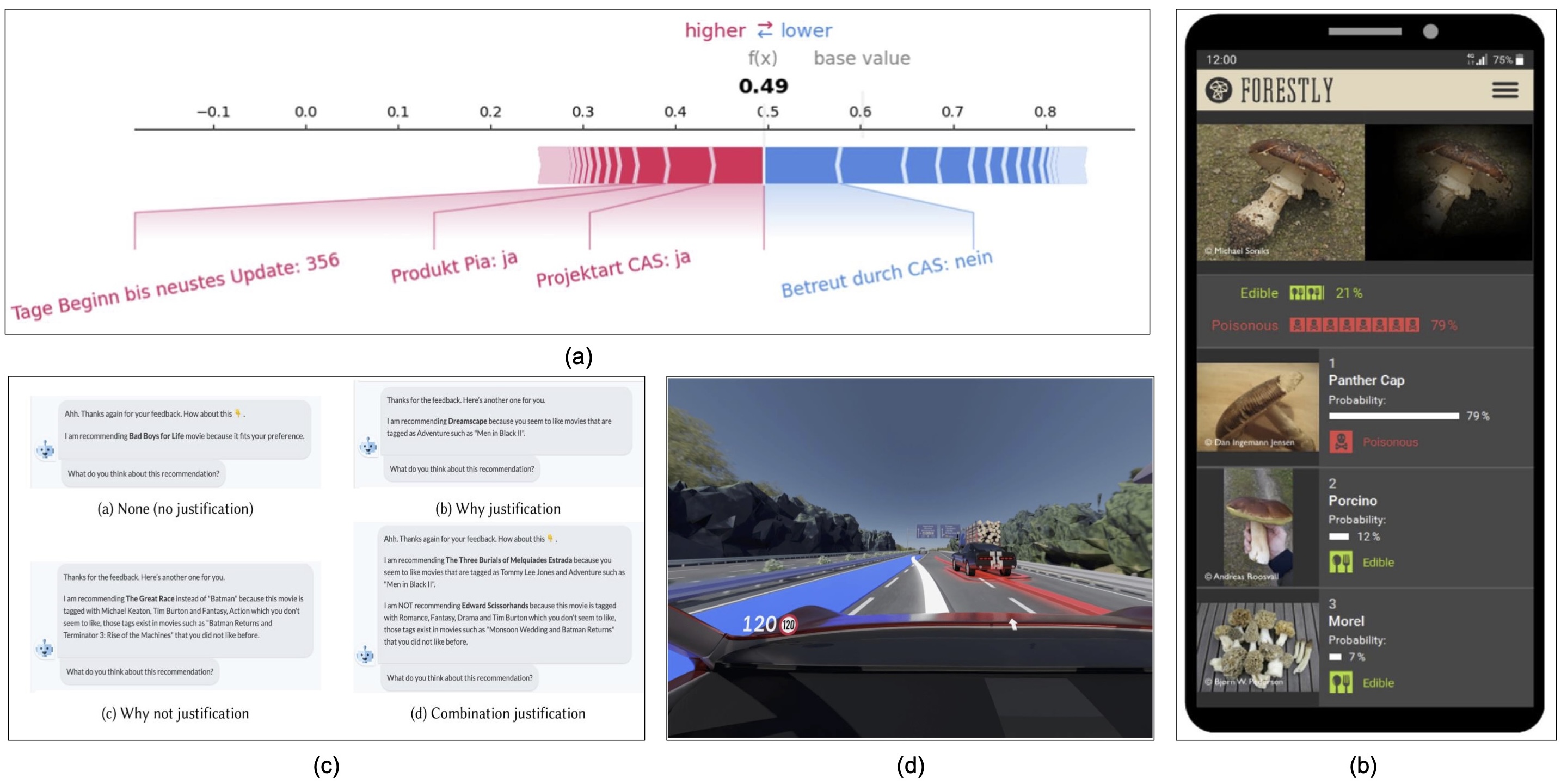}
    \caption{Examples of explainability modalities. Top Row (Left to Right): \cite{schulze-weddige_user_2022,leichtmann_effects_2023}. Second Row: \cite{wilkinson_why_2021,manger_providing_2023}}
    \label{fig:examples-explanations-2}
\end{figure}

\textit{Interaction of interface type and modality} To further understand the pairing between interface types and explanation modalities, we examined how different interactive interface types incorporate various modalities of explanation delivery. Studies using, for example, dashboards, desktop applications, and mobile applications predominantly adopted visual modalities as the basis for delivering explanations, including functional visualizations such as saliency maps, charts, and decision paths (Figure \ref{fig:examples-explanations-2}a). In addition to these functional visual elements, many interfaces also incorporate images, icons, and illustrative graphics to improve visual appeal and assist users in comprehending complex content (Figure \ref{fig:examples-explanations-2}b). In some cases, these graphical elements were supplemented with brief text-based explanations for clarity (Figure \ref{fig:examples-explanations}e). Conversely, chatbots and pop-up messages exclusively relied on text-based explanations (Figure \ref{fig:examples-explanations-2}c). This aligns with the conversational or time-sensitive nature of these interfaces, where explanations are presented in short, sequential, and easily understandable textual formats. These forms of interactions prioritize conciseness and directness, allowing little room for graphical or multi-modal presentations. The single case of Virtual Reality applied graphical-based explanations, embedding explanatory cues directly into the immersive environment (Figure \ref{fig:examples-explanations-2}d). Such visual explanations leveraged the spatial and dynamic affordances of VR, where visual metaphors are more effective than text or audio.

\subsection{Measures of Explainability} \label{subsec:exp_measure}

\subsubsection{Frequency of appeared dimensions}

As outlined in the data analysis section, we mapped the evaluating constructs for explainability into the higher-level categories. Table \ref{tab:distribution} depicts the frequency distribution of the collapsed dimensions within the five main categories. In general, explainability was quantitatively measured 126 times. The most measured category is experiential, accounting for around 23.0\% of the total (N=29). Although usability is the second most measured category (N=26), it encompasses the most dimensions, reflecting its broad and multifaceted nature. The intrinsic categories and the format and presentation are measured 23 and 21 times, respectively. The remaining two categories, ethics and interaction with the explanation, represent a smaller share when measuring explainability. 

For Intrinsic, the most frequently measured dimension is accuracy and fidelity, accounting for 60.9\% of all the dimensions that appeared, followed by quality (34.8\%) and reliability (4.3\%). Format and presentation categories have four dimensions, among which the presence of information is the most frequently measured (42.9\%), followed by the visual aesthetic, modality, and organization of the information. The constructs that examined usability in the included studies have seven unique dimensions. The most frequently measured dimension is understandable, taking up around 38.5\%. Readability follows, making up 15.4\% of the total. The three dimensions--usability, ease of use, and awareness of the explanations--are evenly distributed, each holding 11.5\% of the counts for usability measurements. Finally, both intrusiveness and simulatability were measured twice. 
Five dimensions belong to the experiential category, among which usefulness and satisfaction appeared the most frequently, occurring in more than 82.8\% of the total. The remaining three dimensions--behavioral intention, importance, and interest--appeared in the remaining 17.2\% articles. The ethics category included three key aspects: transparency, trust, and scrutability, which were measured seven, four, and three times, respectively. Four types of user actions are logged and sorted in descending order by measurement proportion: explanation clicks (38.5\%), time spent (30.8\%), access to explanations (23.1\%), and explanation initiation (7.7\%).

\begin{table}
\centering
\caption{Distribution of collapsed dimensions within categories}
\label{tab:distribution}\scalebox{0.55}{
\begin{tabular}{p{2cm}lp{2.5cm}lp{2cm}lp{2.3cm}lp{1.8cm}lp{2.6cm}l}

\toprule 
   \textbf{Experiential}& \textbf{N=29}&
    \textbf{Usability}& \textbf{N=26}&
    \textbf{Intrinsic}& \textbf{N=23}& 
    \raggedright \textbf{Format and Presentation}& \textbf{N=21}&
    \textbf{Ethics}&\textbf{N=14}&
    \raggedright \textbf{Interaction with the Explanation}&\textbf{N=13}\\
    
\midrule 
    Usefulness&14(48.3\%)& Understandable&10(38.5\%)&\raggedright Accuracy and Fidelity & 14(60.9\%)& \raggedright Presence of information &9(42.9\%) & Transparency& 7(50.0\%)&Explanation clicks& 5(38.5\%)\\

  Satisfaction&10(34.5\%)& Readable&4(15.4\%)&Quality & 8(34.8\%)& \raggedright Visual Aesthetics &6(28.6\%)& Trust& 4(28.6\%)&Time spent& 4(30.8\%)\\

  Behavioral intention &3(10.3\%)& Ease of use&3(11.5\%)&Reliability& 1(4.3\%)& \raggedright Modality &4(19.0\%)& Scrutability& 3(23.1\%)& \raggedright Explanation access& 3(23.1\%)\\
	 			
  Importance &1(3.4\%)& Usability&3(11.5\%) &&&\raggedright Organization of the information & 2(9.5\%)& && \raggedright Explanation initiation& 1(7.7\%)\\

 Interest & 1(3.4\%) & \raggedright Awareness of the explanation &3(11.5\%) & && && && & \\
     
 && \raggedright  Intrusiveness&2(7.7\%) & && && && & \\

 &&Simulatability&2(7.7\%)& && && && & \\

\bottomrule 	
\end{tabular}}
\end{table}

\subsubsection{Categories of explainability dimensions}

Table \ref{tab:measurement examples} presents the dimensions of explainability and the categories of dimensions used by the included articles for measuring explainability. It further outlines the synonyms for each dimension as used in the included articles, along with examples of measurement tools and studies. It is worth noting that three studies used an umbrella word (e.g., explainability) to cover their evaluation rubrics without specifying which dimension they measured \cite{moradi_post-hoc_2021,du_leveraging_2022,kou_is_2023}. We did not include the constructs in such studies in our table. 

\begin{longtable}{p{0.13\linewidth}p{0.3\linewidth}p{0.35\linewidth}p{0.18\textwidth}}
\caption{Categories of explainability dimensions}\label{tab:measurement examples}\\

\toprule 
\raggedright \textbf{Categories}& \textbf{Dimensions}&\textbf{Measurement tools examples}& \textbf{Example studies}\\
\midrule
\endfirsthead

\toprule 
\raggedright \textbf{Categories}& \textbf{Dimensions}&\textbf{Measurement tools examples}& \textbf{Example studies}\\
\midrule
\endhead

\midrule
\multicolumn{4}{r}{\textit{(Continued on next page)}}\\
\midrule
\endfoot

\endlastfoot
    
\multicolumn{4}{l}{\textbf{Intrinsic Category}}\\

\raggedright Accuracy and Fidelity 
& \raggedright reasonableness / acceptability / the fidelity of the counterfactual states / perceived component accuracy / suitable & \raggedright  Whether the explanations represented the actual fault.\newline Judging whether a recommended tag is suitable. &  \cite{meas_explainability_2022,wijekoon_user-centred_2023,kenny_explaining_2021,nourani_investigating_2020,olson_counterfactual_2021,yang_explainable_2023} \\

\raggedright Quality &  \raggedright perceived explanation quality/ the quality of explanations & \raggedright System Causability Scale & \cite{schulze-weddige_user_2022, kim_designing_2023,draws_explainable_2023, dikmen_effects_2022}\\

Reliability&  reliability & \raggedright Did you find the explanations to be unpredictable across different queries? & \cite{dey_evaluating_2021} \\

\multicolumn{4}{l}{\textbf{Format and Presentation Category}}\\

\raggedright Visual Aesthetics 
& \raggedright readability of the letters / use of color / use of images 
& \raggedright Rating from 0 to 100 on the readability of the letters / use of color / use of images
& \cite{van_der_waa_evaluating_2021} \\

\raggedright Modality 
& \raggedright enjoyable / preference of different explanation forms
& \raggedright Grade each visualization on a scale from 0 (worst) to 10 (best)
& \cite{tsai_explaining_2019,anjomshoae_context-based_2021,gorski_explainable_2021} \\
\raggedright Organization of information 
& \raggedright organization of information
& \raggedright Rating from 0 to 100 on the organization of information
& \cite{van_der_waa_evaluating_2021} \\

\raggedright Presence of information  & \raggedright completeness / presence of irrelevant, misleading, contradictory information / the information preference in an accompanying explanation & \raggedright This information is complete. \newline I want to be able to view only the most important variables that influence the likelihood of the fault. \newline This explanation has misleading, contradictory, irrelevant information. & \cite{maruf_influence_2023,riveiro_thats_2021,meas_explainability_2022,van2020interpretable} \\

\multicolumn{4}{l}{\textbf{Usability Category}}\\

Understandable 
& understandability / understanding / understandable 
&  This explanation-component is understandable. \newline I understood the warning dialog.
& \cite{zhao_brain-inspired_2023,schoonderwoerd_human-centered_2021,novak_transferring_2022,confalonieri_using_2021} \\

\raggedright Readability 
& \raggedright readable / use of language / familiarity 
& \raggedright The explanation is legible, easy to read.
& \cite{zhao_brain-inspired_2023, van_der_waa_evaluating_2021, desolda_explanations_2023}\\

\raggedright Usability & \raggedright usability & \raggedright It was clear to me how to access the explanations.\newline
The explanation navigation was clear to me.& \cite{silva_explainable_2023,conati_toward_2021,evans_explainability_2022} \\ 

Ease of use & \raggedright perceived ease of use / easiness of interpretation of visual output & \raggedright I would need an expert opinion to understand the explanations provided by SmellChecker.\newline
The color bar / scatter plot / bar plot with the score is easy  to interpret.& \cite{mukhtar_explaining_2023,wysocki_assessing_2023,collenette_explainable_2023} \\

Simulatability & simulatability / user mental model & \raggedright I believe that I could provide an explanation similar to the agent’s explanation.& \cite{silva_explainable_2023,dey_evaluating_2021} \\ 

\raggedright Awareness &  awareness of the explanation & \raggedright Whether they observed the explanation.\newline Whether participants were aware of the provided explanation& \cite{jiang_who_2022,westphal_decision_2023}\\ 

Intrusiveness & intrusiveness / alignment & \raggedright The explanations distracted me from my learning task. \newline This attention map aligns with my attention when reading the natural language prompt.& \cite{conati_toward_2021,kou_is_2023} \\ 

\multicolumn{4}{l}{\textbf{Experiential Category}}\\

\raggedright Satisfaction & satisfaction & \raggedright I was satisfied with the explanations.& \cite{luo_automatic_2023,sreedharan_foundations_2021,symeonidis2022session}\\

\raggedright Usefulness  & \raggedright usefulness / goodness / helpfulness  / sufficiency / value of explanation & \raggedright The explanations provided of how the AI-system works have sufficient detail.\newline The explanations were helpful for me.& \cite{conati_toward_2021,nourani_investigating_2020,neves_interpretable_2021,cruz_evaluating_2022} \\

\raggedright Intention & \raggedright 
willingness to pay for the explanation / intention to use the explanation & \raggedright I want to see this attention when working with code generation models in real life.& \cite{kou_is_2023,kim_how_2023} \\

\raggedright Importance &importance& \raggedright This explanation-component is important.& \cite{schoonderwoerd_human-centered_2021} \\

\raggedright Interest &interest& \raggedright I am not interested in this warning dialog. & \cite{desolda_explanations_2023} \\

\multicolumn{4}{l}{\textbf{Ethics Category}}\\

Transparency & transparency / explanation(s) that best helped them understand a model prediction & \raggedright The visualization presents the relationship between the  recommended person and me.\newline I understand why the cells were marked suspicious through the explanations.& \cite{silva_explainable_2023,tsai_evaluating_2019,mukhtar_explaining_2023}\\ 

Scrutability &  scrutability & \raggedright The visualization allows me to compare and decide  whether the system is correct or wrong.& \cite{balog_transparent_2019,tsai_evaluating_2019} \\

Trust &  trust & \raggedright Did you find the explanations trustworthy for search results? & \cite{dey_evaluating_2021,tsai_evaluating_2019, desolda_explanations_2023} \\

\multicolumn{4}{l}{\textbf{Interaction with the Explanation}}\\

Explanation clicks & Explanation clicks & \raggedright tapping rate of the button\newline number of clicks on the explanation icon & \cite{tsai_explaining_2019,tsai_effects_2021,musto_linked_2019,tsai_evaluating_2019}\\

\raggedright Explanation access & \raggedright Explanation type accessed / Interaction with Explanations  & \raggedright Whether participants accessed the explanation or skipped it. & \cite{naiseh_nudging_2021,conati_toward_2021}\\

Time spent & Time spent on explanation& \raggedright Attention to explanations per hints received \newline Time Spent on Tooltips & \cite{naiseh_nudging_2021,dikmen_effects_2022, conati_toward_2021,dey_evaluating_2021}\\

\raggedright Explanation initiation & Explanation initiation & \raggedright Hints before first explanation initiation \newline Explanation initiations per hints received & \cite{conati_toward_2021}\\

\bottomrule
\end{longtable}

The intrinsic category reflects the inherent attributes of explanations in their own right. The most commonly measured intrinsic feature is \textit{accuracy and fidelity}, which reflects how accurately the explanation mirrors the underlying model. It aims to ensure that the explanation faithfully represents the AI model's internal logic, for example, whether the explanations represent the actual air-handling unit faults \cite{meas_explainability_2022}. \textit{Quality} is a broad term that overlaps with other dimensions to some extent in actual usage. For example, the evaluation of the quality of explanations here is closely tied to usability because a major scale used for measuring it is the system causability scale \cite{schulze-weddige_user_2022,kim_designing_2023,shin_effects_2021}, which was developed based on a previous usability scale and is applied to measure how much the explanation supports users’ causal understanding and how usable the explanation interface is \cite{holzinger_measuring_2020}. \textit{Reliability} refers to the consistency and stability of the explanations provided by a model or system across similar inputs. Only one study examined the reliability of explanations, and it was conducted in the context of searching for fiction books \cite{dey_evaluating_2021}.

The category of format and presentation evaluates the explainable interface to ensure that the explanations are well-organized and represented. Choosing the appropriate \textit{modality} is crucial to maximize the expressive effectiveness of an explanation. An explanation can be delivered to users in different forms: text-based, visual-based, statistical-based, audio-based, or a hybrid of the above. It can be quantified by users’ preference for different explanation presentations \cite{tsai_explaining_2019}. \textit{Aesthetics} concerns whether an interface is visually appealing, e.g., what color it applies to, the appropriate usage of fonts and spacing, or images. Only one study considered visual aesthetics\cite{van_der_waa_evaluating_2021}. The \textit{organization of information} and \textit{presence of information} emphasize how the explanation conveys structured, relevant, necessary information to users. They aim to balance the amount of information according to users' expectations or needs. For example, \cite{meas_explainability_2022} evaluated the importance of different types of information that fault diagnosis explanations might contain, such as the most important variables, probability, etc.

The third category evaluates usability requirements of the explanations to ensure that none of these factors hinder users from utilizing explanations. \textit{Understandable} refers to how easily the target audience can comprehend the explanations. In some cases, \textit{readability} is the foundation of understandability, focusing on the ease with which explanations can be read. \textit{Usability} refers to how easy it is for users to access, navigate, and interact with the explainability content. Though the evaluation of \textit{ease-of-use} overlaps with the above three -- understandability, readability, and usability -- it still deserves to be listed uniquely. Ease of use is sometimes measured by whether users need extra support, i.e., an expert opinion, to understand the explanations. \textit{Simulatability}, which takes a user-centered perspective, measures how well the user is able to recreate or repeat the computational process based on provided explanations of a system \cite{sokol_explainability_2020}. Balanced against the above usability requirements is the need to avoid explanations that attract too much attention from users. On the one hand, it can be checked by manipulating different explanation designs, such as whether users observed the explanation \cite{jiang_who_2022,westphal_decision_2023}. On the other hand, it can be measured as the distraction caused by explanations. For example, Conati et al. included \textit{intrusiveness} in the explanation questionnaire to gauge the potential intrusiveness of the explanations in terms of distraction, confusion, and overwhelming \cite{conati_toward_2021}.   

The fourth category of explainability dimensions focuses on the experiential aspect of the user-explanation interaction. It examines users' experiences, behavioral intentions, and perceptions while engaging with explainability. Experiential measurement is often contextualized by a specific scenario or task.  \textit{Satisfaction} could be measured in different ways. One way was to examine end users’ overall satisfaction with the explanations \cite{mualla_quest_2022, sreedharan_foundations_2021} by using the item, “the explanation of how the simulation tool works is satisfying” \cite{mualla_quest_2022}. In other work, it can be measured in terms of satisfaction with types of visualization techniques \cite{meas_explainability_2022}, satisfaction with explanation styles \cite{symeonidis2022session}, and satisfaction with information provided \cite{tsai_evaluating_2019}, etc. \textit{Usefulness}, \textit{interest}, and \textit{perceived importance} of explanations are also crucial factors when evaluating explanations in intelligent systems. The former is assessed by the end user on how informative, useful, or helpful the explanation is, while the other two relate to whether users are interested in the explanation and whether they perceive the explanation component as important. \textit{Behavioral intention} is often seen as a prerequisite for actual behavior. Unlike other experiential dimensions that measure contextual feelings during the interactive process, \textit{behavioral intentions} focus on evaluating the likelihood of engaging with the explanations in real-world scenarios. For instance, researchers measured whether expert users want to see attention-based explanations when working with code generation models in real life \cite{kou_is_2023}.  

The fifth category, ethics, concerns the alignment of explanations with ethical principles. AI ethics is often broken down into principles in previous studies, including transparency, fairness, responsibility, trust, privacy, etc \cite{vainio2023role, jobin2019global}. In this review, we identify three of them. These three measure ethical considerations of the explainability approach. \textit{Transparency}, considered one of the most prevalent aspects, aims to measure how well an XAI system can explain how the intelligent system works. An explainable approach that achieves transparency should provide explanations covering data, models, and predictions \cite{mcdermid2021artificial}.  \textit{Scrutability} enables users to correct the system's reasoning or modify preferences in the user model \cite{balog_transparent_2019,pu2012evaluating}. For example, in a scrutable recommender system, the user can leverage explanations, e.g., explore and determine the recommendation quality or compare and decide whether the system is correct or wrong \cite{tsai_evaluating_2019}. The final metric, \textit{explanation trustworthiness}, evaluates whether users find the explanation itself trustworthy. This is distinct from a common measure in other studies, where trust refers to whether the explanation helps the user build a stronger sense of trust in the overall system \cite{musto_linked_2019}. 

Unlike the previous five categories, which mainly collect data through self-reported methods such as questionnaires or Likert-scale surveys, the last category focuses on objective metrics related to the participants’ interaction with and various actions related to explanations. These actions include explanation clicks, explanation access, time spent, and explanation initiation. Four studies logged the \textit{explanation clicks} to measure the number of mouse clicks on the explanation icons or to view the explanation interface \cite{tsai_explaining_2019,tsai_effects_2021,tsai_evaluating_2019}. In some tasks, the explanation functionality is triggered by a clickable button, allowing participants to choose whether to enable it. Therefore, \textit{access to explanations} while performing the tasks is logged. \cite{naiseh_nudging_2021} used a binary variable to indicate whether participants accessed the explanation or skipped it . When multiple types of explanations were available to users, the metric was also calculated as a discrete variable to measure explanation types accessed \cite{conati_toward_2021}. The next aspect is the \textit{time spent on explanations}. While other studies track task completion times, the logs for time spent on the explanations can provide insight into user engagement and cognitive effort in the explanation conditions. Two studies utilized an eye tracker to capture how much time the participants spent looking at the explanation \cite{conati_toward_2021,dey_evaluating_2021}. \textit{Explanation initiation},  measured only in one article, indicates how participants initiated explanations in response to hints \cite{conati_toward_2021}.

\section{Discussion} \label{subsec:discussion}

\subsection{Answers to the three research questions}
\subsubsection{Explainability definition}
This literature review identified three key constructs that have recurred in prior explainability research: explainability, explanation, and explainable AI (XAI). Through a textual analysis of these definitions, we identified five dimensions that researchers have used to conceptualize explainability: (1) where explainability is needed, (2) why it is provided, (3) to whom it is directed, (4) what it includes, and (5) how it is conveyed. These dimensions reveal that explainability acts as a conceptual bridge, connecting the content-rich foundation of explanation with the broader objectives and normative expectations associated with XAI. However, our review also uncovered conceptual gaps across these dimensions, which hinder the development of a comprehensive definition of explainability for interactive information systems.

In terms of where explainability is needed, \cite{mcdermid_artificial_2021} proposed a framework identifying the contexts in which explainability can be applied, including data preparation, model development, and explainability visualization. However, our analysis reveals that prior research has predominantly conceptualized explainability from the perspective of model and system development, with limited attention to explainability in data preparation, such as transparency regarding the data used for training algorithms. 

Regarding why explainability is provided, our analysis found that existing research commonly defines explainability through concepts such as transparency, trust, and comprehensibility. These terms embody underlying assumptions about the intended nature of the human–system relationship. For example, when explainability is framed as a way to foster trust, it implies that offering explanations enhances user trust and reliance on the system. Alternatively, when explainability is associated with comprehensibility, the emphasis shifts to ensuring that users gain a thorough understanding of system behaviors and reasoning processes, thereby allowing them to maintain complete control. Future definitions should begin by clarifying the human–system relationship that explainability aims to establish, as the aim shapes both design priorities and evaluation criteria.

As for whom explainability is directed, previous studies suggest that different user groups, such as those with varying levels of AI expertise \cite{mohseni_multidisciplinary_2021}, distinct social roles (e.g., data scientists vs. managers) \cite{barredo_arrieta_explainable_2020}, or domain-specific needs \cite{haque_explainable_2023}, require different types and levels of explainability~\cite{yu2025domain}. This highlights the growing recognition of user diversity, and that explainability is inherently user-oriented and user-sensitive. Yet, existing definitions still employ an overly generic term, such as “user” or "human," overlooking context-specific differences in user needs, perspectives, and capabilities.

Regarding the content included in the explanation, our analysis suggests that previous definitions conceptualize explanation content from several perspectives. The first perspective concerns the granularity of information in explainability. Some scholars adopt a narrow, feature-level approach, defining explainability in terms of the specific features or variables that influence the outputs of AI systems \cite{tesic_can_2022}. In contrast, other scholars propose a much broader view, suggesting that explainability encompasses everything that makes AI systems more understandable to human beings \cite{de_santana_predicting_2023}. These two views reflect fundamentally different assumptions about what types of information are valuable or necessary in human–AI interactions. The second perspective addresses the focus of the explanation, as existing definitions differ in the aspects of the system they aim to make understandable. Some conceptualizations emphasize explanations of the inner workings of models, including the underlying algorithms, decision rules, and reasoning processes \cite{conati_toward_2021,alipour_impact_2020}; whereas other definitions focus on explaining the outputs or predictions of AI systems \cite{tesic_can_2022}, providing users with reasons for current results without revealing the system’s internal mechanisms. The other perspective involves the types of content that serve different functions. For example, causality focuses on the factual relationships between variables \cite{graefe_human_2022}, while insights are more subjective and task-specific, offering guidance for decision-making \cite{wilkinson_why_2021}. This variation suggests that different studies conceptualize explainability with distinct goals in mind, reflecting diverse perspectives on its intended function.

Finally, in terms of how explainability is conveyed, our analysis identified keywords such as ``reveal'' and ``present'' in the literature. While prior studies have extensively discussed the modalities of explanation delivery \cite{bellucci_towards_2021,arya_one_2019}, such as whether the explanation is static or interactive, or whether it is textual or visual, these works primarily answer the operational question of ``how do we explain?'' in terms of the medium or format. In contrast, our analysis of definitions does not refer to the specific form of explanation per se, but rather to describe the intended actions that explanations perform on behalf of the system, whether to ``open up'' the hidden processes, clarify internal reasoning, or depict outputs for the user's understanding. This distinction highlights that definitions of explainability implicitly prescribe the role expected of explanations in shaping human–AI interactions, which is often overlooked in discussions centered on modality.

\subsubsection{Explainability design}
The design of explanation interfaces has gained increasing attention in XAI research, as it fundamentally shapes how real users perceive, interact with, and benefit from explainable systems. While existing reviews have discussed explanation formats and modalities, fewer studies have systematically examined how explanation interfaces are presented in empirical user studies. To address this gap, this paper reviewed interface design practices in explainability research and organized them into three dimensions: interactivity, modality, and interface type.

When it comes to interactivity, while interactive explanation interfaces have gained increasing interest in recent years, a substantial portion of user studies still depend on non-interactive, pre-generated prototypes, such as static screenshots, prepared texts, or mock-ups. This imbalance highlights the technical and resource-related challenges in developing fully functional interactive systems for experimental settings, as well as a reliance on tightly controlled study environments. The existing literature has acknowledged the critical role of interaction affordances in enhancing the effectiveness, transparency, and engagement of explainability interfaces \cite{chromik2021human}. Our findings reaffirm this importance while revealing a methodological inertia that hinders advancing toward user-controllable explanation systems.

Regarding modality, our review proposed a taxonomy for explanation formats, including text, graphical, and voice explanations. It also echoes previous reviews that have emphasized the importance of multimodal explanation designs in addressing diverse user needs and application contexts \cite{vilone2021classification,vilone2020explainable,weitz2021let}. Our review found a clear preference among researchers for text-based and graphical explanations in empirical user studies, while video-based and voice-based explanations remain substantially underexplored. Video-based explanations are intuitive and engaging ways for illustrating dynamic, sequential processes and system behaviors; voice-based explanations hold promise for hands-free, mobile, or accessibility-focused contexts by providing users with real-time, natural language feedback in situations where visual or textual explanations may be impractical. Future studies should actively explore how diverse modalities, especially mixed combinations of modalities, can be leveraged to enhance explanation clarity, thereby better accommodating diverse user needs and application fields.

Our analysis classified deployment platforms for explanation interfaces into nine categories: dashboards, desktop applications, games, chatbots, mobile apps, web-based tools, pop-up messages, virtual reality (VR) systems, and wearable devices. There was uneven attention to these environments. The majority of the included studies relied on graphical user interfaces, particularly dashboards, reflecting their practicality and familiarity in experimental settings. Notably, only a few studies have explored emerging interface types, such as voice user interfaces and virtual reality, despite their significant potential for providing embodied explanation experiences. When examining the pairing patterns between interface types and explanation modalities, our analyses reveal that certain interface types were more likely to be paired with certain modalities. For example, dashboards predominantly employ graphical-based explanations with supplementary text cues, while chatbots and pop-up messages rely almost exclusively on text-based explanations. These pairing patterns suggest an implicit alignment between certain interface types and explanation modalities, likely influenced by technical feasibility, interaction conventions, and user expectations within each environment. Future research should systematically explore how explanation modalities and interface types can be effectively combined to optimize user experience, task performance, and explanation outcomes across diverse interaction scenarios.

\subsubsection{Explainability measurement} Building upon previous classifications \cite{haque_explainable_2023,chromik2020taxonomy}, measurements in explainability studies can be broadly categorized into two types: those assessing the qualities of explanations themselves and those examining their effects. A key motivation for this review lies in the need to clarify the former category in human-centered evaluations of explainable systems. However, a significant limitation in prior work is the insufficient differentiation of individual constructs. For example, several prior studies view understanding as a reflection of how well users comprehend the overall AI-assisted system or task outcome after interacting with explanations, as explanations constitute a type of decision-making assistance \cite{lai2023towards,haque_explainable_2023}. In other cases, however, understanding was measured with respect to the explanation itself, that is, how well users grasp the content, structure, or reasoning process conveyed by the explanation \cite{rong_towards_2023}. By explicitly focusing on the measures of explainability itself, this review clarifies measurement ambiguity and offers a clearer, more structured mapping of what is being measured in user studies on interactive information systems.

A notable finding that emerged from the review is that it provided a comprehensive list of dimensions along which researchers investigate explanations. The Intrinsic category addresses the inherent properties of explanations. Prior research has consistently highlighted the importance of these attributes in explainable systems (e.g., \cite{lofstrom_meta_2022,haque_explainable_2023, lu2023user}), and our review reveals inconsistencies in how they have been empirically measured. For instance, fidelity, which indicates how faithfully an explanation mirrors the underlying model, is sometimes inferred as a factor of trust \cite{hoffman_metrics_2019,jian2000foundations}. Empirical assessments of these attributes, particularly from a user-centered perspective, lack clear operationalization. In contrast, the evaluation technique known as the System Causability Scale is actively gaining traction within the research community \cite{das2020opportunities}. This scale provides a structured and validated framework for assessing the quality of explanations from the user’s viewpoint \cite{holzinger_measuring_2020}. Its growing use reflects a promising step toward standardizing measurements and calls for an effort to develop well-defined instruments for other unspecified dimensions.

The format and presentation category captures how explanations are conveyed to users, and our review identified four primary measurement aspects within this dimension: visual aesthetics, modality, information organization, and information presence. Previous studies have noted the importance of designing and evaluating explainable interfaces \cite{chromik2021human}, and some earlier research has explored individual aspects, for instance, examining modality preferences (e.g., text vs. graph) or types of information presented in the explanations \cite{nguyen_how_2024,kulesza2013too}. However, prior reviews typically treat these as a descriptive characteristic of the system or explanation design, rather than a measurable construct in user evaluations. This highlights a methodological gap in the existing literature, which lacks evaluation of presentation features. In contrast, our review acknowledges format and presentation as an independent, user-centered evaluation focus, documenting how presentation-related constructs have been measured in practice.

Usability has consistently been a central focus in explainability research, with constructs such as understandability, readability, and ease of use widely adopted in user studies. For example, \cite{nunes_systematic_2017} and \cite{rong_towards_2023} both associated ease of use with usability. These measures typically assess users’ subjective perceptions of how easily explanations can be comprehended and utilized within interactive systems. Additionally, we refined usability into measures of awareness and intrusiveness, both important yet underexplored aspects within this category. These focus on whether explanations appropriately catch user attention and whether they interrupt users during their interaction with systems. Balanced against other usability requirements, these two are crucial for ensuring that explanations not only facilitate decision-making but also integrate seamlessly into users’ workflows without causing cognitive overload. 

The ethics category addresses concerns regarding the responsible development and deployment of explainability-assisted systems, focusing on normative attributes such as transparency, scrutability, and trust. These measures emphasize the need to align explanations with broader ethical principles to ensure that users are well-informed, empowered to question decisions, and are able to develop calibrated trust in AI systems. While their conceptual significance is acknowledged, our review highlights that empirical efforts to capture these ethical dimensions systematically remain relatively sparse and inconsistent, indicating an important direction for future explainability research.

The Experiential category refers to users’ subjective experiences and emotional responses when engaging with explanations. Notably, this category emerged as the most frequently measured in our review, underscoring the importance of user experience within the field of explainability research. In line with the previous review \cite{nunes_systematic_2017}, satisfaction and usefulness were measured in 82\% of the articles within this category, indicating their main role as experiential indicators. Conversely, dimensions such as behavioral intention, importance, and situational interest have received less attention, despite their potential to enhance our understanding of how explanations influence users’ engagement, cognitive states, and future behaviors under varying tasks~\cite{liu2019task}. These findings highlight the need for more diverse and theoretically supported experiential measures in explainability evaluations.

Lastly, interaction with the explanation category includes behavioral indicators that reflect how users engage with explanations in interactive information systems. A necessary separation has to be made between measures assessing users’ engagement with the explanations and those evaluating their subsequent outcomes, as the two often do not align. This issue has been largely ignored in previous research, resulting in a methodological gap. For instance, the widely adopted human-AI performance category (e.g.,  Human-AI Task Performance \cite{mohseni_multidisciplinary_2021}) often blends measurements that target different aspects of explainability. Some studies emphasize metrics related to direct interaction with explanations themselves (e.g., \cite{tsai_evaluating_2019,conati_toward_2021}), which reflect how users interact with or rely on explanations during user studies. Others assess whether viewing explanations enhances human performance on tasks, such as acceptance rate or answer accuracy \cite{sardianos_emergence_2021,zhao_brain-inspired_2023}, thereby indirectly inferring the effectiveness of the explanations. While both types of measures are valuable, the lack of differentiation can obscure whether the observed behaviors stem from the explanation’s inherent qualities or from its effects on system use and task outcomes. 

\subsection{Reflections and future directions}
This review not only synthesized the current progress in the explainability of interactive information systems research but also uncovered gaps in the literature. Building on these findings, this section proposes several promising directions to guide future research and advance the research on explainability to support human-AI interaction.

\subsubsection{Advancing the Quality of User Studies}
Our review reveals that while explainability research increasingly adopts more human-centered approaches to assess the effectiveness of explanations, many user studies exhibit limitations in methodological rigor. In particular, issues such as missing data on participant characteristics (e.g., age, gender, and familiarity with AI), unclear research procedures, and unspecified evaluation measures often limit the replicability and generalizability of research findings and re-usability of research resources (e.g. corpora, user dialogue data, annotations and judgment labels)~\cite{liu2022toward}. Given that explainability is intrinsically user/human-centered and its effects usually vary across different user groups and task scenarios, the above gaps might undermine the development of knowledge in the field. To address these limitations, future studies could enhance the quality of user research in two aspects. First, future research should adopt standardized reporting guidelines to ensure the sufficient disclosure of details about user studies. This would enhance the transparency of research processes and enable more systematic comparisons across studies. Second, researchers should create more diverse and theoretically grounded evaluation frameworks that incorporate a broader range of metrics. Furthermore, the explainability research would benefit from the establishment of standardized and validated measurement instruments, similar to the System Causability Scale~\cite{holzinger_measuring_2020}, to promote methodological consistency, robustness in evaluation practices, and effective reuse of research resources (e.g. user interaction data, experimental tasks and hypothetical scenarios, study procedure, prototype interfaces) across different research communities~\cite{jiang2024landscape, liu2021deconstructing}.

\subsubsection{Pairing Definitions with Design and Evaluation Strategies}
A key finding of this review is that there is a disconnect between how explainability is conceptualized, designed, and evaluated in interactive information systems. This observation highlights a central implication of our review: definitions of explainability should serve as the foundation for both system design and evaluation. Specifically, future research should begin by addressing the five dimensions of definition identified in this study — where, why, to whom, what, and how. These five dimensions provide a structured framework for defining the scope and purpose of explainability within a given context.

Once these definitional dimensions are established, the design of explanation interfaces should explicitly align with their corresponding requirements. For example, clarifying where explainability is required (e.g., during data preparation, model output presentation, or decision support) can help determine where explanations should be incorporated into user workflows. Similarly, defining to whom explanations are directed ensures that the style and content of explanations are tailored to the target audience’s knowledge level and needs. The dimensions of why, what, and how can then guide design decisions about the explanation’s intended goals, content, and delivery strategies. Moreover, these same definitional dimensions should also shape the selection of evaluation metrics and research designs. For instance, if a study defines explainability as fostering users’ trust in AI systems, its evaluation should include validated trust scales, as well as behavioral indicators of trust, such as users’ willingness to rely on or reject AI recommendations. Likewise, if explainability is intended to enhance users’ understanding of system reasoning processes, evaluations can employ think-aloud protocols to assess users’ mental models. This definition-driven approach—from definition to design to evaluation—holds great potential for advancing both theoretical insights and practical outcomes in the study of explainable human–AI interaction.

\subsubsection{Explainability in LLM/GenAI Applications}

The rise of large language models (LLMs) and generative AI (GenAI) systems introduces distinctive challenges and opportunities for explainability research~\cite{lubos2024llm}. GenAI is known for various shortcomings, ranging from causing hallucinations in outputs to generating biased or unfair responses. Therefore, explanations can help adjust and validate the output for GenAI \cite{schneider2024explainable}. Previous research has shown that explainability poses greater challenges for GenAI systems due to their open-ended, interactive, and context-sensitive nature, which makes explanations highly dynamic and audience-sensitive \cite{schwalbe2024comprehensive,herrera2025making}. While recent reviews have begun to address explainability in GenAI contexts, such as outlining explainability methods (e.g., local analysis, global analysis) \cite{luo2024understanding}, identifying different stakeholders and deployment contexts for LLMs \cite{meske2022explainable}, and summarizing evaluation measures (e.g., plausibility, truthfulness, consistency) \cite{zhao2024explainability,herrera2025making}, gaps remain in emphasizing the role of audience-centered XAI in LLMs. Building on our findings, we suggest that future research should move beyond system-focused computational explainability methods and prioritize audience-centered approaches for GenAI systems. Such studies will help deepen the understanding of the diverse in-situ and long-term needs, expectations and biases, and potential functional fixedness of various user groups~\cite{liu2025trapped, wang2024understanding, liu2025boundedly}, enabling the development of explanations for GenAI that reflect dynamic sense-making, accommodate varying user expertise, and support situated trust calibration \cite{balayn2025unpacking, corvelo2023human}.

\subsection{Limitations}
Similar to other survey research, this work has some limitations as well as implications for future work. First, as our literature search concluded in February 2024, recent works, particularly those addressing explainability in most recent, domain-specific LLM-based systems, may have been missed. We acknowledge this limitation and strive to incorporate emerging trends into our discussions. Second, while our review systematically examined how explainability is measured, we primarily focused on the measurement of explainability itself. As a result, this survey offers less coverage on evaluating the outcomes or effects of explanations on user perceptions, decisions, and performance. Future research would benefit from a more thorough analysis that combines the evaluation of explanation and its effects on task completion and the fulfillment of user goals. 

\section{Conclusion}
Explainability has become a vital property in the design of interactive information systems, supporting transparency, user trust, and responsible decision-making. To this end, a nuanced understanding of how users interpret and engage with explanations is crucial for advancing both theory and practice in explainable information system design and evaluation measures. Therefore, this review focuses on empirical user studies, systematically examining how explainability has been conceptualized, designed, and measured. 

Following the PRISMA guideline, we synthesize 100 empirical studies and organize the literature according to three aspects: definition, interface design, and evaluation measures of explainability. Together, these contributions offer a consolidated understanding of the current landscape of user-centered explainability research. Conceptually, five dimensions have been identified to capture the multifaceted nature of explainability in interactive information systems. When it comes to design, explanation strategies have been systematically categorized based on their interactive affordances, presentation modalities, and interface type. Evaluation measures are grouped into six user-centered categories, reflecting the diverse ways explainability itself has been assessed in empirical research. Together, these contributions offer a consolidated understanding of the current landscape of user-centered explainability research. This study further confirms three practical avenues for future exploration: enhancing the quality and depth of user studies, systematically aligning definitions with design and evaluation strategies, and developing audience-centered explainability perspectives for LLM and GenAI applications. By pursuing these avenues, future research can assist AI in responsibly integrating into human-centered information environments.

\section*{Acknowledgments}
Special thanks to Yuang-Li Huang for her contribution in analyzing the interfaces. Jiqun Liu's participation in this project is partially supported by a Junior Faculty Summer Fellowship (JFSF) award from the Research Council at the University of Oklahoma.

\bibliographystyle{unsrt}  
\bibliography{main} 

\newpage
\appendix
\section{Appendix}\label{appendix}
\begin{table}[h]
\centering
\scriptsize
\caption{Database and search query}
\label{query}
\begin{tabular}{>{\hspace{0pt}}m{0.063\linewidth}|>{\hspace{0pt}}m{0.142\linewidth}|>{\hspace{0pt}}m{0.137\linewidth}|>{\hspace{0pt}}m{0.515\linewidth}|>{\raggedleft\arraybackslash\hspace{0pt}}m{0.065\linewidth}} \hline
 & Database & Field & Search query & \multicolumn{1}{>{\hspace{0pt}}m{0.065\linewidth}}{Search Results} \\ \hline
Query 1 & The ACM full-text collection & Abstract & "search*" OR "recommendation*" AND
  "user*" AND "explainability" & 77 \\
 & ScienceDirect & Title, abstract or author-specified keywords & ("search" OR "recommendation") AND
  "user" AND "explainability" & 536 \\
 & PsycInfo & Abstract & AB ( "search*" OR "recommendation*" ) AND
  AB ( "user*" AND "explainability*" ) & 11 \\
 & Wiley Online Library & Abstract & ("search*" OR "recommendation*") AND
  "user*" AND "explainability" & 3 \\
 & Web of Science & Abstract & AB=(("search*" OR "recommendation*") AND
  "user*" AND "explainability") & 174 \\
 & Google Scholar & Full text & user explainability search OR recommendation & \multicolumn{1}{>{\hspace{0pt}}m{0.065\linewidth}}{top200} \\
 & IEEE & Title, abstract or author-specified keywords & ("All Metadata":user) AND ("All
  Metadata":explainability) AND ("All Metadata":search) OR
  ("All Metadata":recommendation) & \multicolumn{1}{>{\hspace{0pt}}m{0.065\linewidth}}{top200} \\
 & MIS Quarterly (from Google
  Scholar) & Full text & user explainability search OR recommendation source:MIS
  source:Quarterly & 8 \\
 & IS research (from Google
  Scholar) & Full text & user explainability search OR recommendation
  source:Information source:Systems source: Research & 12 \\
 & Management science (from
  Google Scholar) & Full text & user explainability search OR recommendation source:Management
  source:Science & 23 \\
 & Google Scholar & Full text & explainability AI OR search OR "information
  retrieval" OR recommendation "user evaluation" & \multicolumn{1}{>{\hspace{0pt}}m{0.065\linewidth}}{top200} \\ \hline
Query 2 & The ACM full-text collection & Abstract & “user” AND “explainability” AND (AI OR search OR information
  retrieval OR recommendation) & 147 \\
 & Web of Science & Abstract & “user” AND “explainability” AND (AI OR search OR information
  retrieval OR recommendation) & 234 \\
 & The ACM full-text collection & Abstract & “user” AND “explainability” AND (AI OR search OR information
  retrieval OR recommendation OR conversational agent OR conversational system) & 168 \\
 & Web of Science & Abstract & “user” AND “explainability” AND (AI OR search OR information
  retrieval OR recommendation OR conversational agent OR conversational system) & 232 \\
 & PsycInfo & Abstract & AB (“user” AND “explainability”) AND AB ("AI" OR
  "search" OR "information retrieval" OR
  "recommendation" OR "conversational agent" OR
  "conversational system"), allow relevant terms & 18 \\
 & Wiley Online Library & Abstract & “user” AND “explainability” AND (AI OR search OR information
  retrieval OR recommendation) & 0 \\
 & Wiley Online Library & Abstract & “user” AND “explainability” & 10 \\
 & ScienceDirect & Title, abstract or author-specified keywords & “user” AND “explainability” AND (AI OR search OR information
  retrieval OR recommendation OR conversational agent OR conversational system) & 764 \\
 & Google Scholar & Full text & “user” AND “explainability” AND (AI OR search OR information
  retrieval OR recommendation OR conversational agent OR conversational system) & \multicolumn{1}{>{\hspace{0pt}}m{0.065\linewidth}}{top100} \\
 & IEEE Xplore & Abstract & “user” AND “explainability” AND (AI OR search OR information
  retrieval OR recommendation OR conversational agent OR conversational system) & 95 \\
 & MIS Quarterly (from Google
  Scholar) & Full text & “user” AND “explainability” AND (AI OR search OR information
  retrieval OR recommendation OR conversational agent OR conversational system) & 4 \\ \hline
\end{tabular}
\end{table}

\newpage

\begin{footnotesize}
\begin{longtable}{>{\hspace{0pt}}m{0.16\linewidth}|>{\hspace{0pt}}m{0.11\linewidth}|>{\hspace{0pt}}m{0.65\linewidth}}
\caption{A full list of keyword frequency of the definitions in the three categories}
\label{ap-tab:definition}\\ \hline
\multicolumn{1}{>{\centering\hspace{0pt}}m{0.16\linewidth}|}{\textbf{Keyword}} & \multicolumn{1}{>{\centering\hspace{0pt}}m{0.11\linewidth}|}{\textbf{Frequency}} & \multicolumn{1}{>{\centering\arraybackslash\hspace{0pt}}m{0.65\linewidth}}{\textbf{Example Context}} \endfirsthead \hline
\multicolumn{1}{>{\hspace{0pt}}m{0.16\linewidth}}{\textit{XAI}} & \multicolumn{1}{>{\hspace{0pt}}m{0.11\linewidth}}{~} & ~ \\ \hline
AI & 45 & XAI is a branch of \textbf{AI} intended to explain actions and decisions to humans \\
system & 21 & XAI refers to a \textbf{system} to produce an explanation for the user’s understanding of the model itself and individual decisions of the model \\
understand & 15 & At a high level, XAI is often defined as an initiative or effort made to improve AI transparency and users’ \textbf{understanding} of AI \\
decision & 13 & Explainable Artificial Intelligence (XAI) is a field of study aimed at reliably and efficiently capturing the AI \textbf{decision-making} process, and interpreting and reporting it to a human audience \\
model & 13 & With artificial intelligence dominating in major fields, it becomes imperative to create AI \textbf{models} that are transparent in the sense that the user is presented with an explanation of why the model generated a certain output or made a specific decision, all while preserving the high performance and accuracy of the model \\
user & 13 & Explainable AI (XAI) is a means of AI design where recommendations are supported by explanations to facilitate \textbf{users’} trust calibration process. Explanations provide decisionmakers with insights into how the machine derived a recommendation. Explanations are supposed to help humans identify situations where AI recommendations can be incorrect in specific contexts and cases \\
human & 8 & Explainable Artificial Intelligence (XAI) aims at explaining the algorithmic decisions of AI solutions with non-technical terms in order to make these decisions trusted and easily understandable by \textbf{humans} \\
transparency & 8 & Explainable AI (XAI) is seen as a way to achieve such \textbf{transparency} and interpretability \\
recommendation & 6 & eXplainable AI (XAI) refers to an AI component that explains AI \textbf{recommendations} to humans receiving them. \\
interpretable & 5 & \textbf{Interpretable} machine learning is a subfield of explainable AI, with many different surveys aiming merely to give an overview of inherently interpretable models vs black-box models, while others dive more deeply into the different views and perspectives associated with explainable AI \\
blackbox & 4 & The research field of explainable AI (XAI) tackles the black box problem by introducing transparent models as well as techniques for generating different types of explanations for black box models \\
ML & 4 & explaining the predictions made by \textbf{ML} models \\
specific & 4 & An eXplainable RL (XRL) system that enables humans to correctly understand the agent’s aptitude in a \textbf{specific} task \\
trust & 4 & Explainable AI (XAI) is a means of AI design where recommendations are supported by explanations to facilitate users’ trust calibration process \\
help & 3 & It helps people understand the decision-making process of AI algorithms by bringing transparency and accountability into AI systems \\
prediction & 3 & Explainable AI (XAI) has emerged as a field to address this need for AI systems’ \textbf{predictions} to be followed by explanations of these \textbf{predictions} \\
result & 3 & XAI can be defined as the field that aims to make AI systems \textbf{results} more understandable to humans \\
technology & 3 & Explainable Artificial Intelligence (XAI) aims at explaining the algorithmic decisions of AI solutions with \textbf{non-technical} terms in order to make these decisions trusted and easily understandable by humans \\
inner workings & 3 & Explainable AI (XAI) suggests that having AI systems explain their \textbf{inner workings} to their end users \\
accuracy & 2 & With artificial intelligence dominating in major fields, it becomes imperative to create AI models that are transparent in the sense that the user is presented with an explanation of why the model generated a certain output or made a specific decision, all while preserving the high performance and \textbf{accuracy} of the model \\
algorithm & 2 & Explainable Artificial Intelligence (XAI) aims at explaining the \textbf{algorithmic} decisions of AI solutions with non-technical terms in order to make these decisions trusted and easily understandable by humans \\
improve & 2 & At a high level, XAI is often defined as an initiative or effort made to \textbf{improve} AI transparency and users’ understanding of AI (Adadi and Berrada, 2018). \\
output & 2 & A major goal of Explainable Artificial Intelligence (XAI) is to have AI-systems construct explanations for their own \textbf{output}. \\
reason & 2 & Explainable models attempt to provide \textbf{reason} and causality behind their decisions \\
reveal & 2 & explainability in AI (XAI) as methods that \textbf{reveal} how AI decisions are formed, helping users to understand and appropriately trust AI systems \\ \hline
\multicolumn{1}{>{\hspace{0pt}}m{0.15\linewidth}}{\textit{Explanation}} & \multicolumn{1}{>{\hspace{0pt}}m{0.11\linewidth}}{~} & ~ \\ \hline
decision & 10 & A useful explanation model would help users to understand the recommendation reasoning process, which allows the users to make a better \textbf{decision} or persuade them to accept the suggestions from a system \\
recommendation & 10 & Explanations are generally initiated by an information provider to clarify or justify the \textbf{recommendation} and convince the recipient to comply with it \\
model & 7 & Explanation: an interface between human and system that accurately approximates the \textbf{model} of the system and is comprehensible to the human \\
understand & 7 & how the reasoning process of an intelligent information system should be made \textbf{understandable} to the user \\
user & 7 & We consider explanations in the framework of counterfactual reasoning, where a \textbf{user} who is confused by the agent’s activity (or proposed activity) presents alternative behavior that they would have expected the agent to execute. \\
reason & 6 & An explanation could be “a summary of some \textbf{reasoning} or provenance for facts” \\
feature & 5 & Explanations may simply offer insight into how the decision was reached (e.g., highlighting important \textbf{features}, presenting answer probabilities, etc.) \\
help & 5 & An explanation is sometimes a justification of why items have been recommended, while sometimes an item description which \textbf{helps} users to understand the qualities of the item well enough to decide whether it is relevant for them or not \\
item & 4 & Generally, explanations “seek to show how a recommended \textbf{item} relates to a user’s preferences” \\
system & 4 & Explanation: an interface between human and \textbf{system} that accurately approximates the model of the system and is comprehensible to the human \\
describe & 3 & a contrastive explanation \textbf{describes} "why event A occurred as opposed to some alternative event B" \\
human & 3 & “explanation”, a term that in this context denotes any indication that could help the \textbf{human} decision-maker (in our case, the ECG reader) understand the output of the decision support (in our case an ML classifier) \\
information & 3 & Explanations are generally initiated by an \textbf{information} provider to clarify or justify the recommendation and convince the recipient to comply with it \\
insight & 3 & One popular type of explanation, the justification, offers \textbf{insight} as to why a decision is a good one without necessarily describing the algorithmic details \\
justification & 3 & An explanation is sometimes a \textbf{justification} of why items have been recommended, while sometimes an item description which helps users to understand the qualities of the item well enough to decide whether it is relevant for them or not \\
present & 3 & A thorough explanation should explain what imaging features are \textbf{present} in those important locations, and how changing such features modifies the classification decision \\
alternative & 2 & We consider explanations in the framework of counterfactual reasoning, where a user who is confused by the agent’s activity (or proposed activity) presents \textbf{alternative} behavior that they would have expected the agent to execute \\
answer & 2 & in an informal way, an explanation can be seen as a narrative \textbf{answering} the question of why particular facts (e.g., an event or a decision) are happening. Such a narrative may explain things in terms of causality \\
behavior & 2 & a global explanation where the common \textbf{behaviours} of the tool are explained \\
causality & 2 & Explanations aim to reach a high level of \textbf{causability}, which can be defined as the extent to which users understand an explanation in an efficient, effective, and satisfying manner \\
classification & 2 & A thorough explanation should explain what imaging features are present in those important locations, and how changing such features modifies the \textbf{classification} decision \\
quality & 2 & An explanation is sometimes a justification of why items have been recommended, while sometimes an item description which helps users to understand the \textbf{qualities} of the item well enough to decide whether it is relevant for them or not \\
relate & 2 & Generally, explanations “seek to show how a recommended item \textbf{relates} to a user’s preferences” \\
specific & 2 & a local explanation where feature attribution explains why a \textbf{specific} code snippet is predicted to be vulnerable \\ \hline
\multicolumn{1}{>{\hspace{0pt}}m{0.16\linewidth}}{\textit{Explainability}} & \multicolumn{1}{>{\hspace{0pt}}m{0.11\linewidth}}{~} &  \\ \hline
human & 9 & explainability requires an accurate description of a decision-maker and defines how comprehensible the decisions are to \textbf{humans} \\
model & 9 & In a broader view, explainability encompasses everything that makes ML \textbf{models} transparent and understandable, also including information about the data, performance, etc \\
system & 9 & explainability emphasizes making ML \textbf{systems} comprehensible to humans through explanations \\
understand & 9 & Explainability is crucial for users to \textbf{understand} why certain results are being presented to them, particularly in the case of health information where the consequences of acting on incorrect or misleading information can be severe \\
user & 8 & Explainability focuses on making the recommendation process and the reasons behind specific recommendation more clear to the \textbf{users} \\
ability & 7 & Explainability includes the \textbf{ability} of humans to understand the explanation \\
information & 7 & Explainability is the term applied to the concept whereby a user is provided with sufficient \textbf{information} to be able to reconstruct why an AI-driven system made a prediction \\
decision & 6 & explainability methods that aid humans in verifying models’ \textbf{decisions} or help them make better \textbf{decisions} \\
AI & 5 & Explainability has been identified as a requirement to promote reliability and trust in the \textbf{AI} output and also to ensure humans remain in control \\
item & 4 & the system’s ability to be able to explain to users why specific \textbf{items} are recommended \\
recommendation & 4 & explainability, i.e., the system may just need to justify why the \textbf{recommendation} was presented \\
xxx-based & 3 & Common approaches to (post hoc) explainability of specific predictions of AI systems include feature importance, saliency maps, and \textbf{example-based} methods \\
interpretable & 3 & \textbf{Interpretability} is a related term and can be defined as the level to which the user understands and can make use of the explanations given by the system and the information provided \\
ML & 3 & Explainability emerged as a research area aiming to address interpretability bottlenecks in \textbf{ML} models resulting from the fact that many of them can be effective in predicting an outcome, but not in explaining their underlying reasoning, which limits their practical application in critical areas \\
output & 3 & Explainability has been identified as a requirement to promote reliability and trust in the AI \textbf{output} and also to ensure humans remain in control \\
prediction & 3 & Explainability is the term applied to the concept whereby a user is provided with sufficient information to be able to reconstruct why an AI-driven system made a \textbf{prediction} \\
specific & 3 & the system’s ability to be able to explain to users why \textbf{specific} items are recommended \\
clearer & 2 & the ability of a model to make its functioning \textbf{clearer} to an audience \\
comprehensible & 2 & explainability requires an accurate description of a decision-maker and defines how \textbf{comprehensible} the decisions are to humans \\
everything & 2 & Explainability or Explainable AI (XAI) can be defined as \textbf{everything} that makes AI more understandable to human beings. \\
feature & 2 & Common approaches to (post hoc) explainability of specific predictions of AI systems include \textbf{feature} importance, saliency maps, and example-based methods \\
functioning & 2 & the ability of a model to make its \textbf{functioning} clearer to an audience \\
internal & 2 & Explainability is the extent to which the \textbf{internal} mechanics of algorithmic journalism can be explained in understandable human language \\
need & 2 & explainability, i.e., the system may just \textbf{need} to justify why the recommendation was presented \\
present & 2 & Explainability is crucial for users to understand why certain results are being \textbf{presented} to them, particularly in the case of health information where the consequences of acting on incorrect or misleading information can be severe \\
reason & 2 & Explainability focuses on making the recommendation process and the \textbf{reasons} behind specific recommendation more clear to the users \\ \hline
\end{longtable}
\end{footnotesize}

\end{document}